\documentclass[letterpaper]{article} 
\usepackage{aaai24}  
\usepackage{times}  
\usepackage{helvet}  
\usepackage{courier}  
\usepackage[hyphens]{url}  
\usepackage{graphicx} 
\usepackage{natbib}  
\usepackage{caption} 
\usepackage{booktabs}
\usepackage{multirow}
\usepackage{algorithm}
\usepackage{algorithmic}
\usepackage{subcaption}
\usepackage{enumitem}
\usepackage{amsmath}
\usepackage{amssymb}
\usepackage{mathtools}
\usepackage{amsthm}
\usepackage{placeins}
\usepackage{thmtools} 
\usepackage{thm-restate}
\usepackage[T1]{fontenc}
\theoremstyle{plain}
\newtheorem{theorem}{Theorem}[section]

\newtheorem{lemma}[theorem]{Lemma}
\newtheorem{corollary}[theorem]{Corollary}
\theoremstyle{definition}

\newtheorem{assumption}[theorem]{Assumption}
\theoremstyle{remark}

\usepackage{newfloat}
\usepackage{listings}
\DeclareCaptionStyle{ruled}{labelfont=normalfont,labelsep=colon,strut=off} 
\floatstyle{ruled}
\newfloat{listing}{tb}{lst}{}
\floatname{listing}{Listing}
\title{Algorithmic Decision-Making under Agents with Persistent
Improvement}
\author{
    Tian Xie\textsuperscript{\rm 1},
    Xuwei Tan\textsuperscript{\rm 1},
    Xueru Zhang\textsuperscript{\rm 1}
}
\affiliations{
    \textsuperscript{\rm 1}Department of Computer Science and Engineering, the Ohio State University\\
    \{xie.1379, tan.1206, zhang.12807\}@osu.edu\\
}

\usepackage{bibentry}

\begin{document}

\maketitle
\begin{abstract}
This paper studies algorithmic decision-making under human strategic behavior, where a decision maker uses an algorithm to make decisions about human agents, and the latter with information about the algorithm may exert effort strategically and improve to receive favorable decisions. Unlike prior works that assume agents benefit from their efforts immediately, we consider realistic scenarios where the impacts of these efforts are persistent and agents benefit from efforts by making improvements gradually. 
We first develop a dynamic model to characterize persistent improvements and based on this construct a Stackelberg game to model the interplay between agents and the decision-maker. We analytically characterize the equilibrium strategies and identify conditions under which agents have incentives to invest efforts to improve their qualifications. With the dynamics, we then study how the decision-maker can design an optimal policy to incentivize the largest improvements inside the agent population. We also extend the model to settings where 1) agents may be dishonest and game the algorithm into making favorable but erroneous decisions; 2) honest efforts are forgettable and not sufficient to guarantee persistent improvements. With the extended models, we further examine conditions under which agents prefer honest efforts over dishonest behavior and the impacts of forgettable efforts. 
\end{abstract}

\section{Introduction}\label{sec: intro}

In applications such as lending, college admission, hiring, recommendation systems, etc., machine learning (ML) algorithms have been increasingly used to evaluate and make decisions about human agents. Given information about an algorithm, agents subject to ML decisions may behave strategically to receive favorable decisions. How to characterize the strategic interplay between algorithmic decisions and agents, and analyze the impacts they each have on the other, are of great importance but challenging

This paper studies algorithmic decision-making under strategic agent behavior. Specifically, we consider
a decision-maker who assesses a group of agents and aims to accept those that
are \textit{qualified} for certain tasks based on assessment outcomes. With knowledge of the acceptance rule, agents
may behave strategically to increase their chances of getting accepted. For example, agents may invest to genuinely \textit{improve} their qualifications (i.e., honest effort), or they may  \textit{manipulate} the observable assessment outcomes to game the algorithm (i.e., dishonest effort). Both types of behaviors have been studied. In particular, \citet{Hardt2016a,Dong2018,Braverman2020,Hardt2021,Sundaram2021,zhang2022,Eilat2022} focus on learning under strategic manipulation, where they 
proposed various analytical frameworks (e.g., Stackelberg games) to model manipulative behavior, and analyzed models or developed learning algorithms that are robust against manipulation.  

Another line of research \citep{zhang2020, harris2021stateful,bechavod2022information,Kleinberg2020, Chen2020,Barsotti2022,Jin2022} considers a different
setting where agent qualifications (labels) change in accordance with the improvement actions. The
goal of the decision-maker is to design a mechanism such
that agents are incentivized to behave toward directions that improve the underlying qualifications. Notably, \citet{Kleinberg2020} proposed a mechanism to incentivize individuals to invest in specific improvable features. Their work inherited the classical settings of the Principal-agent model in economics but designed an incentivizing mechanism under a linear classifier. They modeled manipulation and improvement similarly (linear in efforts) and did not consider the persistent and delayed effects of improvement. The mixture of both improvement and manipulation behavior is also studied \citep{Miller2020,Chen2020, Barsotti2022, horowitz2023causal,Jin2022}. However, these works regarded improvement as a similar action to manipulation where the only difference is it will incur a label change. Another related topic is \textit{performative prediction} \citep{perdomo2020, Izzo2021, hardt2022p,jin2024performative}, an abstraction that captures agent actions via model-induced distribution shifts.  
Details and more related works are presented in Sec. \ref{sec:related}.


This paper primarily focuses on honest agents (i.e., agents will invest effort to improve their qualifications), while settings with both improvement and manipulation are also studied. We first propose a novel two-stage Stackelberg game to model the interactions between decision-maker and agents, i.e., the decision-maker commits to its policy, following which agents best respond. A crucial difference between this study and the prior works is that the existing models all assume that the results of agents' improvement actions are \textit{immediate}, i.e., once agents decide to improve, they experience \textit{sudden} changes in qualifications and receive the return \textit{at once}. However, we observe that in many real-world applications, the impacts of improvement action are indeed \textit{persistent} and \textit{delayed}. For example, humans improve their abilities by acquiring new knowledge, but they make progress gradually and benefit from such behavior throughout their lifetime; loan applicants improve their credit behaviors by repaying all the debt in time, but there is a time lag between such behaviors and the increase in their credit scores. Therefore, it is critical to capture these delayed outcomes in the Stackelberg game formulation.

To this end, we propose a \textit{qualification dynamic} model to characterize how agent qualifications would gradually improve upon exerting honest efforts. Such dynamics further indicate the time it takes for agents to reach the targeted qualifications that are just enough for them to be accepted. The impacts of such time lag on agents are then captured by a \textit{discounted utility model}, i.e., reward an agent receives from the acceptance diminishes as time lag increases. Under this discounted utility model, agents best respond by determining how much effort to exert that maximizes their discounted utilities.  

This paper aims to analytically and empirically study the proposed model. With the understanding of the strategic interactions between the decision-maker and agents, we further study how the decision-maker can design an optimal policy to incentivize the largest improvements inside the agent population, and empirically verify the benefits of the optimal policy.

Additionally, we extend the model to more complex settings where (i) agents have an additional option of strategic manipulation and can exert dishonest effort to game the algorithm; (ii) honest efforts exerted by agents are forgettable and may not be sufficient to guarantee persistent improvements, instead the qualifications may deteriorate back to the initial states. We will propose a \textit{model with both manipulation \& improvement} and a \textit{forgetting mechanism} to study these settings, respectively. We aim to examine how  agents would behave when they have both options of manipulation and improvement, under what conditions they prefer improvement over manipulation, and how the forgetting mechanism
affects an agent’s behavior and long-term qualifications. 

Our contributions can be summarized as follows:
\begin{enumerate}[leftmargin=*,topsep=-0.1em,itemsep=-0.1em]
    \item We formulate a new  Stackelberg game to model the interactions between decision-maker and strategic agents. To the best of our knowledge, this is the first work capturing the delayed and persistent impacts of agents' improvement behavior (Sec.~\ref{sec:problem}).
    \item We study the impacts of acceptance policy and the external environment on agents, and identify conditions under which agents have incentives to exert honest efforts. This provides guidance on designing incentive mechanisms to encourage agents to improve (Sec.~\ref{sec:improvement}). 
    \item We characterize the optimal policy for the decision-maker that incentivizes the agents to improve (Sec.~\ref{sec:educator}).
    \item We consider the possibility of dishonest behavior and propose a \textit{model with both improvement and manipulation}; we identify conditions when agents prefer one behavior over the other (Sec.~\ref{sec:manipulate}).
    \item We propose a \textit{forgetting mechanism} to examine what happens when honest efforts are not sufficient to guarantee persistent improvement (Sec.~\ref{sec:forget}).
    \item We conduct experiments on real-world data to evaluate the analytical model and  results (Sec.~\ref{sec:exp}). 
\end{enumerate}

\section{Related Work}\label{sec:related}
\subsection{Strategic Manipulation}\label{app:manip}

Though our work primarily lies in proposing a new model for improvement behaviors, the problem settings are also closely related to strategic classification problems \cite{Hardt2016a,Ben2017,Dong2018,Braverman2020,Sundaram2021,Hardt2021,ahmadi2021strategic,Eilat2022,horowitz2023causal,miehling2019strategic}. \citet{Hardt2016a} formulated classification problems with strategic manipulation as a Stackelberg game with deterministic cost functions, where the decision maker optimizes classification accuracy based on individuals' best responses. Afterwards, more sophisticated analytical frameworks were proposed \cite{Dong2018, Braverman2020, Hardt2021}. \citet{Dong2018} proposed an online algorithm for strategic classification. \citet{Braverman2020} added randomness to strategic classifiers, while \citet{shao2024strategic, lechner2023strategic} provided a complete analysis of the regret bound for online strategic classification. On the other hand, \citet{Sundaram2021} analyzes the statistical learnability of strategic classification with an SVC classifier. \citet{Hardt2021} relaxed the \textit{standard microfoundations} assumption where individuals are perfectly rational to \textit{alternative microfoundations} where a proportion of individuals may not be strategic, and proposed a \textit{noisy response model} to tackle the new problem. \citet{zhang2022} studied the setting where the decision maker and individuals only have knowledge of the feature distributions as random variables. Thus, the strategic manipulation corresponds to a distribution shift and its cost is also a random variable. \citet{Eilat2022} considered the setting where individual responses are dependent and the classifier is learned through \textit{graph neural networks}. \citet{xie2024automating} studied the long-term impacts on welfare and fairness under a sequential strategic learning setting.

\subsection{Improvement}\label{app:improve}

However, there are other literature considering improvement behavior \cite{Lydia2019,Rose2020,shavit2020,Alon2020,zhang2020,Chen2020,Kleinberg2020,bechavod2021,ahmadi2022,ahmadi2022setting,raab2021unintended,xie2024learning, xie2024non,zhang2020long}. Unlike strategic manipulation, improvement will incur a label change. \citet{Lydia2019} studied the conditions where fairness interventions can promote improvement among individuals. \citet{zhang2020} formulated the label change as a transition matrix where the transition probabilities are deterministic and difficult to estimate. Other works consider both behaviors at the same time. \citet{xie2024learning} studied randomness when the agents can both improve and manipulate, while \citet{xie2024non} relaxed the linear assumption on the decision policy and studied the welfare under strategic learning settings. \citet{Kleinberg2020} proposed a mechanism to incentivize individuals to invest in specific features where the individuals have a budget to invest strategically on all features including undesired ones. Their work inherited the classical settings of the Principal-agent model in economics but designed an incentivizing mechanism under a linear machine learning classifier. They modeled manipulation and improvement similarly (linear in efforts) and did not consider the persistent and delayed effects of improvement. By contrast, we first develop a fundamentally different dynamic model to characterize persistent and delayed improvements. Based on the new model, we construct a Stackelberg game to model the interplay between agents and the decision-maker. 

Another line of literature focuses on the underlying causal models of strategic classification. \citet{shavit2020} and \citet{Alon2020} introduced causal inference frameworks into strategic behaviors including manipulation and improvement. \citet{Chen2020} divided the features into immutable features, improvable features and manipulable features and explored linear classifiers which can prevent manipulation and encourage improvement. \citet{Jin2022} also focused on incentivizing improvement and proposed a subsidy mechanism to induce improvement actions and improve social well-being metrics. \citet{Barsotti2022} conducted several experiments where both improvement and manipulation are possible and both actions incur linear deterministic costs. 

\subsection{Other Agent Dynamics }\label{app:recomm}
In addition to manipulation and improvement, some studies consider different individual behaviors and models. For example, \citet{zhang2019group,dean2024emergent,hashimoto2018fairness} focused on participation dynamics where agents at each time decide their participation in decision systems. \citet{perdomo2020, hardt2022p, jin2024addressing} studied \textit{performative prediction} which is an abstract framework for optimization when the deployed model influences agent population. \citet{yin2024long} studied the setting where the population distribution has an unknown dynamic and reinforcement learning is used to improve long-term fairness. We refer interested readers to survey papers \cite{zhang2021fairness,dean2024accounting} for more examples.
 Also, our work is related to preference shifts and opinion dynamics in recommendation systems, which we refer to \citet{castellano2009statistical} as a comprehensive survey. Among the rich set of works, \citet{Dean2022, gaitonde2021polarization} proposed geometric models for opinion polarization and motivate our work. 

\section{Problem Formulation}\label{sec:problem}

Consider an agent population with $m$ skill sets. Each agent has a \textit{qualification profile} at time $t$, denoted as a unit $m$-dimensional vector $q_t\in [0,1]^{m}$ with $||q_t||_2=1$, 
A decision-maker at each time makes decisions $D_t\in\{0,1\}$ (``0" being reject and ``1" accept) about the agents based on their qualification profiles. Let fixed vector $d\in [0,1]^m$ be the ideal qualification profile that the decision maker desires.  

\paragraph{Decision-maker's policy.} For an agent with qualification profile $q_t$, the decision-maker assesses whether the agent's profile lines up with the desired qualifications $d$, and makes decision $D_t$ based on their similarity $x_t:=q_t^Td$ using a \textit{fixed} threshold policy $\pi(x_t) = \textbf{1}(x_t\geq \theta)$, i.e., only agents that are sufficiently fit can get accepted. How to choose threshold $\theta$ is discussed in Sec.~\ref{sec:educator}. We assume only agents with initial similarity $x_0\geq 0$ are interested in positions and only focus on these candidates.   

Although the decision policy introduced above focuses on the similarity between $q_t$ and $d$ where qualifications $q_t$  are normalized with the same magnitude for all agents, it can be easily extended to settings where the magnitude/strength of skills also matters and may differ across agents. For example, if the decision-maker prefers students with balanced math and English skills and there are two "balanced" students, the decision-maker is likely to prefer the one with higher scores. Therefore, we propose a \textbf{pre-normalization procedure} to take magnitude into account. The idea is to first add an additional dimension to initial qualification profile $q_0$, which represents the agent's unobservable ``irrelevant attribute" (all other skills an agent has that are not important for the decision). Thus, we expand the original $q_0$ to obtain a $m+1$ dimensional complete qualification profile. Meanwhile, we add a dimension to the ideal qualification profile $d$ with $0$ as its value; the new ideal profile becomes $[d; 0]$. Then we can make the following natural assumption:

\begin{assumption}\label{assumption: prenorm}
After adding the dimension of “irrelevant attribute”, for all agents, the norms of their complete qualification profiles are the same. 
\end{assumption}

Assumption \ref{assumption: prenorm} has been supported by literature in machine learning \cite{liu22} and social science \cite{holmstrom1991multitask}. The ``irrelevant" dimension demonstrates all other skills that belong to an agent but are not important to the decision. Therefore, competency in relevant/measurable attributes implies weakness in irrelevant/immeasurable attributes and the length of the complete qualification profile stays the same for all agents. With Assumption \ref{assumption: prenorm} and the distribution of $q_0$ as $Q$, we formalize the \textbf{pre-normalization} procedure in Algorithm \ref{alg:main}.

\begin{algorithm}
  \caption{Pre-normalization procedure}
  \label{alg:main}
  \begin{algorithmic}[1]
    \REQUIRE Joint distribution $Q$ for $q_0$, n agents with $\{q_0^i\}_{i=1}^{i=n}$ where $q_0^i \in [0,1]^{m}$, $d \in [0,1]^{m} $.
    \ENSURE Normalized $\{q_0^i\}_{i=1}^{i=n}$ (i.e., $q_0^i \in [0,1]^{m}$ and $\|q_0^i\| = 1$), new $d \in [0,1]^{m+1}$.
    \STATE $d = [d,0]$.
    \STATE According to $Q$, find the largest norm $K = \max_{q_0 \sim Q}\lVert q_{0}^i \rVert$ of original profiles.
    \FOR{$i \in \{1,\ldots n\}$}
        \STATE Calculate norm difference $z^i = \sqrt{K^2 - \lVert q_{0}^i \rVert_2^2}$.
        \STATE $q_0^i = \frac{[q_{0}^i; z^i]^T}{K}\in[0,1]^{m+1}$.
    \ENDFOR
  \end{algorithmic}
\end{algorithm}

\paragraph{Agent qualification dynamics.} We assume agents have information about the ideal profile $d$ (e.g., from application guides, mock interviews). In the beginning, agents with $q_0$ can choose to improve their profiles by investing an effort $k\in[0,1]$ to acquire the relevant knowledge, but the effort will have delayed and persistent effects over subsequent time stages.  The specific value of $k$ depends on the agent's utility and will be introduced at the end of this section. Upon investing $k$, the agent's qualifications $q_t$ gradually improve over time based on the following:

\begin{eqnarray}\label{eq:dynamics}
\widetilde q_{t+1} = q_t + k\cdot q_t^T d\cdot d~; ~~~~~~~~  q_{t+1} = \frac{\widetilde q_{t+1}}{\| \widetilde q_{t+1} \|_2}. 
\end{eqnarray}

\begin{figure}
    \centering
    \includegraphics[width=0.2\textwidth]{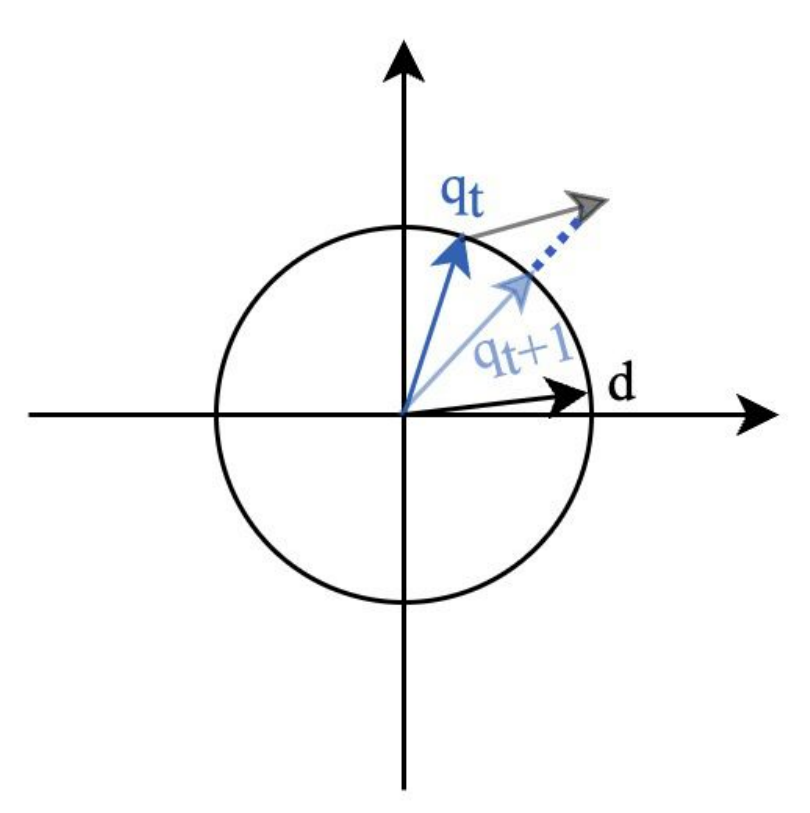}
    \caption{{Dynamics of agent's qualification $q_t$.}}
    \label{fig:dynamics}
\end{figure}

\eqref{eq:dynamics} suggests that agents at each time improve toward the ideal profile $d$. How much they can improve depend on their current profile $q_t$ and the effort $k$. The similarity $q_t^Td$ in the dynamics captures the reinforcing effects:  agents that are more qualified could have more resources and are more capable of leveraging the acquired knowledge to improve their skills. Note that the maximum improvement an agent attains at each round is bounded, i.e., the normalized vector $q_{t+1}$ after improvement is always between current qualifications $q_t$ and the ideal profile $d$. Fig.~\ref{fig:dynamics} illustrates the improvement dynamics of qualification $q_t$ in a two-dimensional space.

Dynamics in~\eqref{eq:dynamics} model the delayed and persistent impacts of improvement action (i.e., effort $k$). In many real applications, humans acquire knowledge and benefit from repeated practices. They make progress toward the goal gradually, and it takes time to receive the desired outcome from the investment. Indeed, \eqref{eq:dynamics} is inspired by the dynamics in \citet{Dean2022}, which was used for modeling preference shifts (details are in Sec.~\ref{app:recomm}), where individuals update their opinions/preferences based on their correlations with some influencer (e.g., a political figure) and control the power of the intervention. We believe this is similar to improvement especially when agents improve themselves by imitating some "role models". For example, in a job application scenario, the “influencer” is a current worker who holds information session and introduces her profile $d$. Then the agents strive to mimic the profile of $d$ by updating $q_t$. The imitating nature of improvement is well justified in many works (e.g., \citet{raab2021unintended, zhang2022}), making agent improvement suitable to be modeled in a similar fashion to concept drift/preference shift. Thus, we use \eqref{eq:dynamics} to model the evolution of agents' (pre-normalized) qualifications. Based on Prop. 1 of \citet{Dean2022}, we know that $q_t$ converges under dynamics, as formally stated in Lemma \ref{lemma:convergence} below.

\begin{lemma}[Convergence of qualification]\label{lemma:convergence}
Consider an agent with initial similarity $x_0:=q_0^Td > 0$. If he/she makes an effort $k$ and improves qualification profile $q_t$ based on dynamics in~\eqref{eq:dynamics}, then $q_t$ converges to the desired profile $d$. 
The evolution of the similarity $x_t:=q_t^Td$ is given by: 
\begin{eqnarray}\label{eq:convergence}
x_{t}^{-2} - 1 = \frac{(x_0)^{-2}-1}{(k+1)^{2t}}
\end{eqnarray}
\end{lemma}
Lemma~\ref{lemma:convergence} suggests that any agent eventually becomes an ideal candidate with a perfectly aligned profile (i.e., $x_t = q_t^Td = 1$), as long as he/she is interested in the position ($x_0\geq 0$) and willing to make an effort ($k>0$). The only difference among agents is the speed of convergence: it takes less time for agents who are more qualified at the beginning (i.e., larger $x_0$) and/or make more effort  (i.e., larger $k$) to become ideal and get accepted. Note that our work focuses on agent's improvement behavior with \textit{persistent} and \textit{delayed} effects. Although the model is presented in a simplified setting where only a one-step effort is made by the agents at the beginning, it can capture more complicated scenarios where agents repeatedly exert efforts multiple times until they reach the target. Each effort has persistent effects on improving the qualification. Specifically, suppose each agent at time $t$ can exert an effort $k_t$ and the agent's qualification improves based on $\widetilde q_{t+1} = q_t + \sum_{\tau=0}^t k_\tau\cdot q_t^T d\cdot d$ with $q_{t+1} = \frac{\widetilde q_{t+1}}{\| \widetilde q_{t+1} \|_2}$ (i.e., every time the agent improves from all accumulated efforts $\sum_{\tau=0}^t k_\tau\in[0,1]$ he/she invested so far). For this new dynamics, the overall impacts of these efforts $\{k_t\}_{t\geq0}$ on improving agent qualification can be \textit{equivalently} characterized by the dynamics~\eqref{eq:dynamics} with some one-step effort. That is, there exists an effort $k^*\in[0,1]$ such that investing $k^*$ once at the beginning has the same impact on $\lim_{t\to \infty}q_t$ as investing a sequence of efforts $\{k_t\}_{t\geq0}$ over time. We provide more detailed discussion on this in App.~\ref{app:relax}. We also discuss a special case where the effect of $k$ is decreasing over time and provide further convergence analysis (Thm. \ref{theorem:generalized}) in App.~\ref{app:relax}.

\paragraph{Agent's utility \& action.} Because it takes time for agents to receive rewards (i.e., get accepted) for their efforts, they may not have incentives to invest if there is a long delay. In practice, people may be more attracted to investments with immediate rewards than delayed rewards, or they may simply not have enough time to wait. For example, students only have limited time to prepare for college applications;  credit card applicants may not have incentives to improve their credit scores and wait to get approval for a specific credit card when there are many instant-approval cards on the market.

To characterize the delayed rewards, we use a discount model and assume the reward each agent receives from the effort $k$ decreases over time. Specifically, let $H$ be the minimum time it takes for an agent to get accepted from the effort $k>0$. We define \textbf{agent's utility} as:  
\begin{eqnarray}\label{eq:utility1}
U = \frac{1}{(1+r)^{H}} - k.
\end{eqnarray}
That is, the utility is the exponentially discounted reward an agent receives from the acceptance minus the effort. $r>0$ is the discounting factor.  Note that the discounted utility model\footnote{Under exponential discounting function, the agent's reward diminishes at a constant rate \citep{grune2015models}. Our model can also adopt other discounting functions (e.g., hyperbolic discounting) for settings when the agent's reward decreases inconsistently. The qualitative results of this paper still remain the same.} has been widely used in literature such as reinforcement learning \citep{kaelbling1996}, finance \citep{meier2013}, and economics \citep{krahn1993, samuelson1937note}. 

Since threshold policy $\pi(x_t)=\textbf{1}(x_t\geq \theta)$ is used to make decisions, an agent gets accepted whenever the qualification profile is sufficiently aligned with the ideal profile, i.e.,  $x_t=q_t^Td\geq \theta$. Based on ~\eqref{eq:convergence}, we can derive $H$ as a function of threshold $\theta$, agent's initial similarity $x_0$, and effort  $k$, i.e.,
\begin{eqnarray}\label{eq:H}
H &=& \min_{t}~\{x_t \geq \theta\} = \min_{t}~\left\{\frac{(x_0)^{-2}-1}{(k+1)^{2t}} \leq \frac{1}{\theta^2}-1 \right\} \nonumber \\
&=& \frac{-\ln\left(\sqrt{\frac{(\theta)^{-2}-1}{(x_0)^{-2}-1}}\right)}{\ln(k+1)}
\end{eqnarray}
Plug in \eqref{eq:utility1}, agent's utility becomes: 
\begin{eqnarray}\label{eq:utility}
U:= U(k,\theta,r,x_0) = {(1+r)^{\frac{\ln\left(\sqrt{\frac{(\theta)^{-2}-1}{(x_0)^{-2}-1}}\right)}{\ln(k+1)}}} - k.
\end{eqnarray}
Therefore, strategic agents will choose to improve their qualifications only if utility $U(k,\theta,r,x_0)>0$, and they will choose the investment $k$ that maximizes the utility.






\paragraph{Stackelberg game.} We model the strategic interplay between the decision-maker and agents as a Stackelberg game, which consists of two stages: (i) the decision-maker first publishes the optimal acceptance threshold $\theta$ (details are in Sec.~\ref{sec:educator}); (ii) agents after observing the threshold take actions to maximize their utilities as given in \eqref{eq:utility}.

\paragraph{Manipulation \& forgetting.} The model formulated above has two implicit assumptions: (i) agents are honest and they improve their qualifications by making actual efforts; (ii) once agents make a one-time effort $k$ to acquire the knowledge, they never forget and can repeatedly leverage this knowledge to improve their profiles based on \eqref{eq:dynamics}. However, these assumptions may not hold. In practice, agents may fool the decision-maker by directly manipulating $x_t$ to get accepted without improving actual   $q_t$, e.g., people cheat on exams or interviews to get accepted. Moreover, the knowledge agents acquired at the beginning may not be sufficient to ensure repeated improvements.  To capture these, we further extend the above model to two settings: 

\begin{enumerate}[leftmargin=*,topsep=0.1em,itemsep=0.1em]
    \item \textit{Manipulation:} Besides improving the actual profile $q_t$ by making an effort $k$, agents may choose to manipulate $x_t$ directly to fool the decision-maker. The detailed model and analysis are in Sec.~\ref{sec:manipulate}.
    \item \textit{Forgetting:} One-time investment $k$ may not guarantee the improvements all the time, i.e., qualifications $q_t$ do not always move toward the direction of ideal profile $d$, instead it may devolve and possibly go back to starting state $q_0$. The detailed model and analysis are in Sec.~\ref{sec:forget}.
\end{enumerate}

\paragraph{Objective.} In this paper, we study the above interactions between decision-maker and agents. We aim to understand (i) under what conditions agents have incentives to improve their qualifications; (ii) how to design the optimal policy to incentivize the largest improvements inside the agent population; (iii) how the agents would behave when they have both options of manipulation and improvement, and under what conditions agents prefer improvement over manipulation; (iv) how the forgetting mechanism affects agent's behavior and long-term qualifications.


\section{Improvement \& Optimal Effort}\label{sec:improvement}

In this section, we examine the impact of decision threshold $\theta$ and the environment (i.e., discounting factor $r$) on agent behavior. Specifically, we focus on agents with discounted utility (\eqref{eq:utility}) and identify conditions under which the agents have incentives to improve their qualifications. Note that we do not consider issues of manipulation and forgetting in this section. Based on \eqref{eq:utility}, an agent with  $x_0:=q_0^Td$ chooses to improve only if its utility $U(k,\theta,r,x_0) > 0$. To characterize the impact of an agent's one-time investment $k$ on $U(k,\theta,r,x_0)$, we first define a function $C(\theta, r, x_0)$ that summarizes the impacts of all the other factors (i.e., threshold $\theta$,  discounting factor $r$, and initial profile similarity $x_0$) on agent utility, as defined below.
\begin{eqnarray}
C(\theta, r, x_0) = -\ln\left(\sqrt{\frac{(\theta)^{-2}-1}{(x_0)^{-2}-1}}\right)\cdot \ln(1+r)
\end{eqnarray}
Based on $C(\theta, r, x_0)$, we can derive conditions  under which agents have incentives to improve (Thm.~\ref{theorem: effort}).

\begin{theorem}[Improvement \& optimal effort]\label{theorem: effort}
There exists a threshold $m> 0$ such that for any $\theta, r, x_0$ that satisfies $C(\theta, r, x_0) < m$,
the agent has the incentive to improve the qualifications, i.e., agent utility is positive for some efforts $k>0$. 
 Moreover, there exists a unique optimal effort $k^* \in (0,1)$ that maximizes the agent utility. 
\end{theorem}

Thm.~\ref{theorem: effort} identifies a condition under which agents have incentives to exert positive effort $k>0$. This condition depends on factors $\theta,r,x_0$ and can be fully characterized by the function $C:=C(\theta,r,x_0)$. 

\begin{figure}[h]
\includegraphics[trim=0cm 0.0cm 0.7cm 0.4cm,clip,width=0.75\columnwidth]{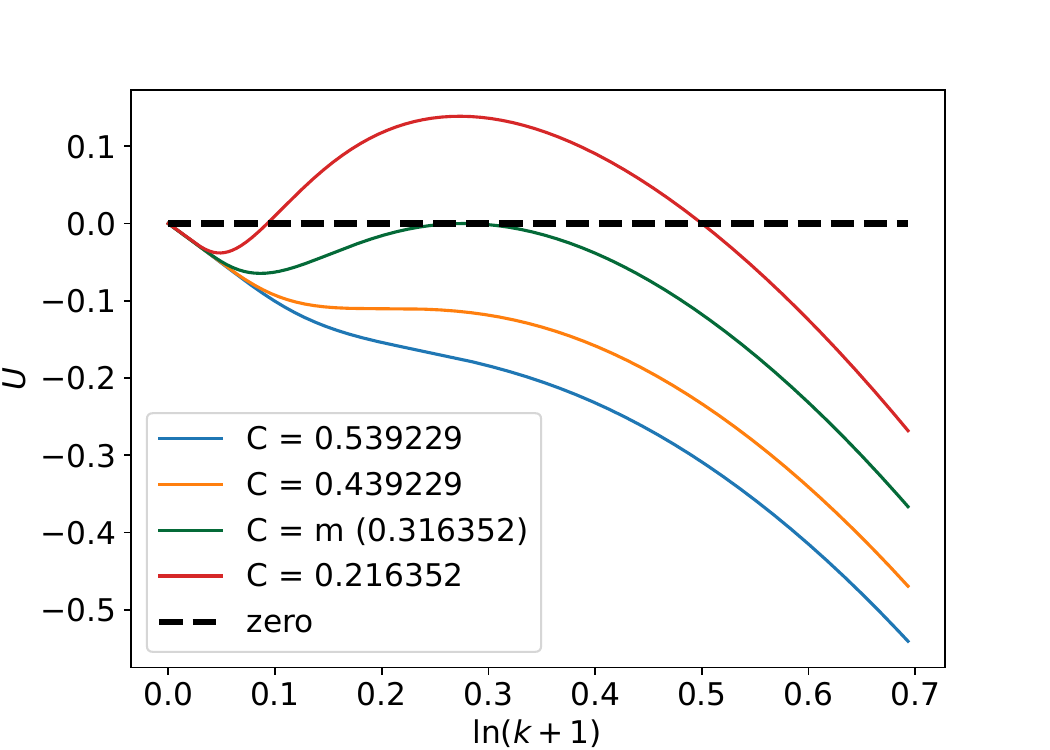}
\centering
\caption{Impact of effort $k$ on agent utility $U$ under different $C:=C(\theta,r,x_0)$: $\exists m>0$ such that agents have incentives to invest and improve their qualifications if $C< m$.}
\label{fig: improvement utility}
\vspace{-0.2cm}
\end{figure}

Although the analytical solution of the threshold $m$ is difficult to find, we can numerically solve $m\approx 0.3164$ as shown in App. \ref{subsection: proof1}. In Fig.~\ref{fig: improvement utility}, we illustrate agent utilities $U$ as functions of effort under different $C$. The results show that only when $C<m$ (red curve), an agent can attain positive utility with effort $k>0$; when $C\geq m$ (green/yellow/blue curve), agents will not invest  because the maximum utility is attained at $k=0$. Moreover, when $C<m$ (red curve), there is  a unique optimal effort $k$ that maximizes the utility. These results are consistent with Thm.~\ref{theorem: effort}.

The condition in Thm.~\ref{theorem: effort}  further indicates the impacts of policy $\theta$, discounting factor $r$, and initial state $x_0$ on agent behavior. Specifically, agents only invest if $C(\theta, x_0, r)<m$ holds. By fixing any two of $\theta, x_0, r$, we can identify the domain of the third factor under which agents invest to improve. These results are summarized in Table~\ref{table: domain} and verified in App.~\ref{app:corollary3}. It shows that for any threshold $\theta$ and discounting factor $r$, agents only improve if their initial qualification profile is sufficiently similar to the ideal profile; the domain of $\theta$ also implies the best profile an agent with initial state $x_0$ can reach after exerting effort: if acceptance threshold $\theta$ is larger than the upper bound of $\theta$ given in Table~\ref{table: domain}, then agents will not have incentives to improve.

\begin{table}[h]
\begin{center}
\caption{Domain of initial similarity $x_0$ (or threshold $\theta$) under which agents invest positive efforts.
}
\label{table: domain}
\vspace{-0.2cm}
\renewcommand{\arraystretch}{1.05}
\resizebox{0.45\textwidth}{!}{
\begin{tabular}{ccc}\toprule
\textbf{Domain of $x_0$ (given $\theta,r$)} & $x_0 > \left(1+ (\theta^{-2} - 1)\cdot \exp\left(\frac{2m}{\ln(1+r)}\right)\right)^{-1/2}$   \\
\midrule
\textbf{Domain of $\theta$ (given $x_0,r$)} & $\theta \le \left(1 + (x_0^{-2} - 1)\cdot \exp\left(-\frac{2m}{\ln(1+r)}\right)\right)^{-1/2}$ \\ 
\bottomrule
\end{tabular}
}
\end{center}
\end{table}
\vspace{-0.3cm}

The above results further suggest effective strategies that encourage agents to improve their qualifications, i.e., more agents are incentivized to improve if (i) the decision-maker's acceptance threshold $\theta$ is lower; or (ii) the time it takes for agents to succeed after investments is shorter (smaller discounting factor $r$). Examples of both strategies in real applications are discussed in App.~\ref{app:corollary3}, which further verify the effectiveness of our proposed model.





\section{Decision-maker's policy to incentivize improvement}\label{sec:educator}

Sec.~\ref{sec:improvement} studied the impact of threshold $\theta$ on agent behavior and provided guidance on incentivizing agents to improve. In practice, although it is more difficult to adjust the discounting factor $r$, the decision-maker can adjust the threshold policy $\theta$ to incentivize the largest possible amount of total improvement, thereby improving the \textit{social welfare}. In this section, we study the optimal policy when the decision-maker is aware of the agent's best response and hopes to incentivize agents to improve.

Suppose the decision-maker has full information about agents and can anticipate their behaviors, i.e., for any decision threshold $\theta$, it knows that agents whose initial similarity $x_0>x^*(\theta):=\left(1+ (\theta^{-2}-1)\cdot \exp\left(\frac{2m}{\ln(1+r)}\right)\right)^{-\frac{1}{2}}$ will invest and improve their profiles (by Table~\ref{table: domain}). Also, we define $x^*(0) = 0$ to let $x^*(\theta)$ be continuous in $[0,1]$ and denote $f$ as the probability density function of the agent similarity $x_0$ which is also continuous in $[0,1]$. Then, we can define $U_d(\theta)$ as the utility of the decision-maker under the threshold as the total amount of agents' improvements:
\begin{eqnarray}\label{eq: Ud}
    U_d(\theta) = \int_{x^*(\theta)}^{\theta} (\theta-x_0)\cdot f(x_0) dx_0 
\end{eqnarray}

Eq.~\eqref{eq: Ud} above demonstrates that the decision-maker aims to maximize the total improvement among the agent population, and its utility is a function of $\theta$. Since $f(x), x^*(\theta)$ are both continuous in $[0,1]$, utility $U_d(\theta)$ is also continuous. The following Thm.~\ref{theorem: optimal theta} further shows the existence of the optimal thresholds $\theta^* \in (0,1)$.

\begin{theorem}[Existence of optimal threshold]\label{theorem: optimal theta}
For any decision-maker with utility function $U_d$, there exists at least one $\theta^* \in (0,1)$ that is optimal under which $U_d(\theta)>0$. Moreover, $\theta^*$ is the unique optimal point of $U_d$ if $\frac{\partial U_d}{\partial \theta}$ has one root within $(0,1)$.
\end{theorem}

\begin{figure}[h]
     \centering
 \begin{subfigure}[b]{0.3\columnwidth}
     \centering        \includegraphics[width=\linewidth]{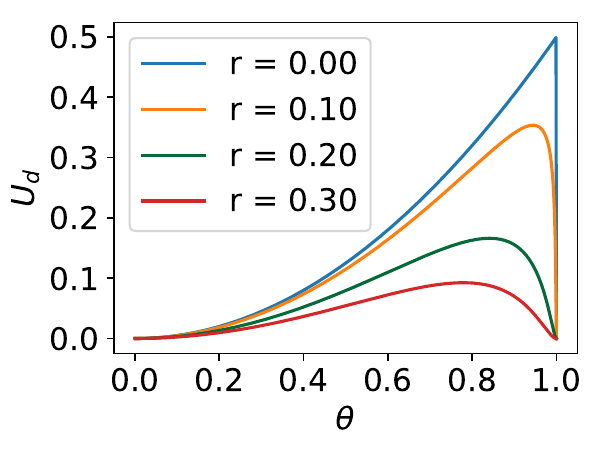}
         \caption{$f = U(0,1)$}
     \end{subfigure}
     \hspace{0.5em}
     \begin{subfigure}[b]{0.3\columnwidth}
     \centering       \includegraphics[width=\linewidth]{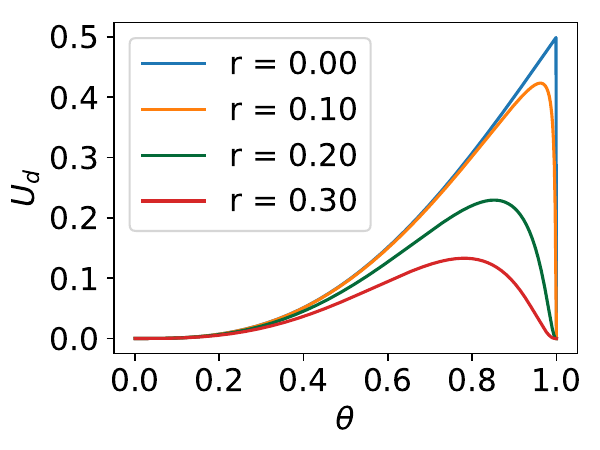}
         \caption{$f = Beta(2,2)$}
     \end{subfigure}
     \hspace{0.5em}
     \begin{subfigure}[b]{0.3\columnwidth}
     \centering       \includegraphics[width=\linewidth]{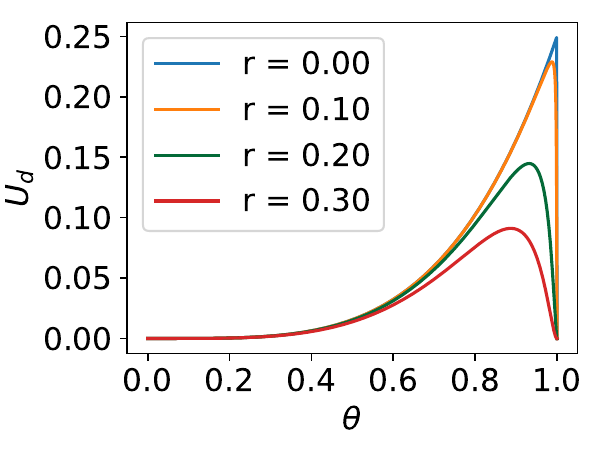}
         \caption{$f = Beta(3,1)$}
     \end{subfigure}
     \caption{Optimal thresholds $\theta^*$ under different density functions $f$ and discounting factors $r$.}\label{fig:illuedu}
     \centering
     \vspace{-0.3cm
     }
\end{figure}

To verify Thm.~\ref{theorem: optimal theta}, we demonstrate the values of $U_d$ under situations where the agent population has different density functions $f$ and different discounting factors $r$. Specifically, we consider the uniform distribution and Beta distributions with different parameters. Fig.~\ref{fig:illuedu} shows $U_d(\theta)$ under different density functions $f$ and discounting factors $r$. The results illustrate that under these settings, $U_d$ is single-peaked and there is a unique $\theta^*\in(0,1)$ that is optimal and results in positive utility, which is consistent with Thm.~\ref{theorem: optimal theta}. The figure also indicates the impact of  $r$ on the optimal threshold: as $r$ increases, $\theta^*$ increases and the corresponding maximum utility decreases. As formally stated below in Corollary \ref{corollary: theta change vs r}. We prove Thm. \ref{theorem: optimal theta} and Corollary \ref{corollary: theta change vs r} in App. \ref{subsection:proof1.5}.

\begin{corollary}\label{corollary: theta change vs r}
For $U_d(\theta)$ that has a unique maximizer $\theta^*$,  optimal $\theta^*$ decreases as $r$ increases.
\end{corollary}
Importantly, the results of Thm.~\ref{theorem: optimal theta} show that the decision-maker can always find an optimal decision threshold $\theta^*$ (either numerically or using gradient methods depending on the density function $f$) to incentivize the largest improvement and promote \textit{social welfare} in practice. While the above results all assume the decision-maker knows $r$ when determining $\theta$, we can relax this and provide a procedure to estimate $r$; this is included in App. \ref{app:restimate}.

\section{Impact of Manipulative Behavior}\label{sec:manipulate}

Our analysis and results so far rely on an implicit assumption that agents are honest and they improve qualifications $q_t$ by making actual efforts. However, as mentioned in Sec.~\ref{sec:problem}, agents in practice may fool the decision-maker by strategically manipulating $x_t=q_t^Td$ to get accepted without improving $q_t$. Next, we extend our model in Sec.~\ref{sec:problem} by considering the possibility of such  manipulative behavior.

\paragraph{Model with both manipulation \& improvement.} We extend the model in Sec.~\ref{sec:problem} where agents after observing $\theta$ have an additional option to manipulate $x_0$ directly.  If they choose to \textbf{\textit{improve}}, they make a one-time effort $k\in[0,1]$ to acquire relevant knowledge and gradually improve their qualifications $q_t$ over time based on \eqref{eq:dynamics}. If they choose to \textit{\textbf{manipulate}}, they only increase $x_t$ at every round to fool the decision-maker without changing the actual profile $q_t$. Similar to the literature on strategic classification \citep{Hardt2016a}, the manipulation comes at the cost  and the risk of being caught.

Specifically, let $c(x',x)\geq 0$ be the \textit{manipulation cost} it takes for an agent to increase its similarity  from $x$ to $x'$, and $P\in[0,1]$ be the  \textit{detection probability} of manipulation during an agent's entire application process. Agents, once getting caught manipulating $x_t$, will never be accepted.  

\paragraph{Degree of manipulation.} If agents choose to manipulate, they will increase $x_t$ at every round to fool the decision-maker, and they manipulate in a way that minimizes the manipulation cost and the risk of being detected. We make the following natural assumptions on $c$ and $P$:\\
\begin{enumerate}[leftmargin=*,topsep=-0.1cm,itemsep=-0.cm]
    \item  Let $\overline{x}_{t}$ be the best outcome agents can attain from $x_{t-1}$ at round $t$ by improvement behavior (with largest effort $k=1$). If $x_t>\overline{x}_{t}$ for some $t$, then $P=1$ because the decision-maker can be certain that $x_t$ is the result of manipulation; otherwise, $P\in[0,1)$ if $x_t\leq \overline{x}_{t}$.
    
    \item The total manipulation cost it takes for an agent with initial similarity $x_0$ to be accepted is $c(\theta,x_0)$.\\
\end{enumerate}

Note that $\overline{x}_t$ above indicates the maximum degree of manipulation of agents: to avoid being detected, an agent should not manipulate $x_t$ more than $\overline{x}_t$. We can compute $\overline{x}_t$ directly from Lemma~\ref{lemma:convergence} (by setting $k=1$), i.e.,  
$\overline{x}_t= \left(\frac{x_{t-1}^{-2}-1}{4}+1\right)^{-\frac{1}{2}}$.
For agents who manipulate, if the total manipulation cost needed to get accepted is $c(\theta,x_0)$ and  detection probability $P=1$ whenever $x_t>\overline{x}_t$, then agents will always manipulate toward  $\overline{x}_t$ to maximize its utility. As a result, agents who manipulate can be regarded as they mimic the improvement behavior with the largest effort $k=1$.

Let $\widetilde{U}$ be agent's \textbf{utility under manipulation}, which  is the benefit an agent obtains from acceptance (when not being detected) minus the manipulation cost, i.e., 
\begin{eqnarray}\label{eq: utilde}
\widetilde{U} =(1-P)\cdot (1+r)^\frac{-\ln\left(\sqrt{\frac{(\theta)^{-2}-1}{(x_0)^{-2}-1}}\right)}{\ln2} - c(\theta, x_0),
\end{eqnarray}
where the benefit is derived based on \eqref{eq:utility} (with $k=1$).

\paragraph{Agent's best response.} Suppose agents have full information about detection probability $P$ and discounting factor $r$, after observing the acceptance threshold $\theta$, they best respond by choosing the action (i.e.,  improvement/manipulation/do nothing) that maximizes their utilities, i.e., if $\widetilde{U}>\max_k {U}$, they choose to  manipulate; otherwise, they  improve by exerting optimal effort $k^*=\arg\max_k {U}$.

Next, we examine under what conditions agents prefer improvement over manipulation.

\begin{theorem}\label{theorem: choice}
Suppose manipulation cost $c(x',x) = (x' - x)_+$ and threshold $\theta \ge \bar{\theta}$ for some $\bar{\theta}\in(0,1)$. For any discounting factor $r$, there exists $ \widehat{P}\in(0,1)$ such that the followings hold:

1. If $P = 0$, then $\exists \widehat{x}\in(0,1)$ such that agents manipulate only when initial similarity $x_0 \in (\widehat{x},\theta)$.

2. If  $P \in (0, \widehat{P}]$, then $\exists \widehat{x_1},\widehat{x_2}$ such that agents manipulate only when initial $x_0 \in (\widehat{x_1}, \widehat{x_2})$.

3. If  $P > \widehat{P}$, then agents never choose to manipulate.
\end{theorem}

Thm.~\ref{theorem: choice} considers scenarios when the threshold is sufficiently high, and  identifies conditions under which manipulation is preferred by agents in these settings. It shows agent behavior highly depends on the risk of manipulation (i.e., detection probability $P$). 
 The specific values of  $\widehat{P}$, $\widehat{x}$, $\widehat{x_1}$, $\widehat{x_2}$ in Thm.~\ref{theorem: choice} depend on $\theta$, $r$. In particular, $\widehat{P}$ increases as $r$ increases. 
Indeed, we can empirically find  $\widehat{P}$, $\widehat{x}$, $\widehat{x_1}$, $\widehat{x_2}$ and verify the theorem. These are illustrated in App. \ref{app:thm5}  and Sec.~\ref{sec:exp}.

\section{Forgetting Mechanism}\label{sec:forget}
The analysis in previous sections relies on the assumption that once agents
make a one-time effort $k$ to acquire the knowledge, they
never forget and can repeatedly leverage this knowledge to
improve their profiles based on \eqref{eq:dynamics}. This may not hold in practice when the knowledge
agents acquired at the beginning are not sufficient
to guarantee repeated improvements.
In this section, we extend the qualification dynamics (\eqref{eq:dynamics}) by incorporating the \textit{forgetting mechanism}, i.e., qualification profile $q_t$ does
not always move toward the direction of ideal profile $d$,
instead, it may devolve and possibly go back to the initial $q_0$. Note that we only consider honest agents who do not manipulate.  By modifying  \eqref{eq:dynamics}, we define the new \textbf{qualification dynamics with forgetting} as follows. 
\begin{eqnarray}\label{eq:fdynamic}
\widetilde q_{t+1} &=& q_t +  (k\cdot d + (1-k)\cdot q_0)\cdot q_t^Td\\
\nonumber q_{t+1} &=& \frac{\widetilde q_{t+1}}{\|\widetilde q_{t+1}\|_2}
\end{eqnarray}
Let $\widetilde{d} := k\cdot d + (1-k)\cdot q_0$, then new dynamics in~\eqref{eq:fdynamic} implies that at each round, qualification profile $q_t$ is pushed toward the direction of $\widetilde{d}$, i.e., a convex combination of  ideal profile $d$ and initial qualifications $q_0$. Whether $q_t$ improves  towards $d$ or deteriorates back to $q_0$ depends on the investment $k$: with more effort $k$, the degree of forgetting is less; there is no forgetting  if all the knowledge is acquired ($k=1$). Under the new dynamics, we can derive the convergence of the qualification profile as follows.

\begin{theorem}[Convergence of qualification under forgetting]\label{theorem: fconverge}
Consider an agent with initial similarity $x_0=q_0^Td>0$ whose qualifications $q_t$ follow dynamics in~\eqref{eq:fdynamic}. Suppose the agent makes investment $k>0$, then $q_t$ converges to profile $d^*$ and the similarity $x_t=q_t^Td$ satisfies:
\begin{eqnarray}\label{eq: fconverge}
(x_{t}^*)^{-2} - 1 < \frac{(x_0^*)^{-2}-1}{(k_u+1)^{2t}}
\end{eqnarray}
where $d^* =  \frac{\widetilde{d}}{\|\widetilde{d}\|}$,
 $x_t^* = q_t^Td^*$, and $k_u = \| \widetilde{d} \|\cdot x_0$. 
\end{theorem}

\begin{figure}
    \centering
     \includegraphics[width=0.6\linewidth]{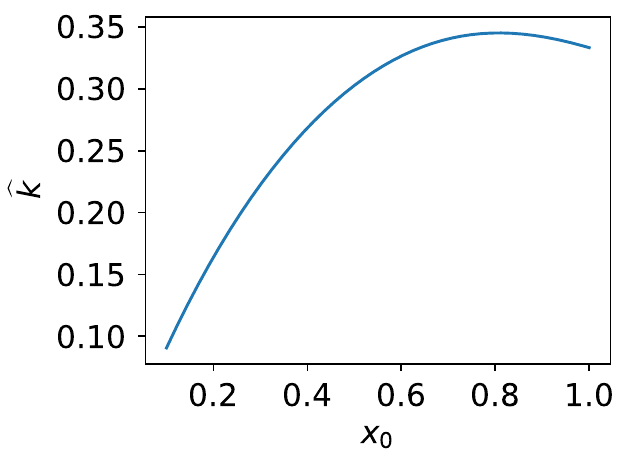}
    \caption{Upper bound $\widehat{k}$ of the optimal effort  as a function of $x_0$. }
    \label{fig:khat}
\end{figure}



Thm.~\ref{theorem: fconverge} implies that convergence still holds when qualifications evolve with forgetting. Unlike the scenarios without forgetting where $q_t$ eventually converges to the ideal profile $d$ regardless of  $k$ (Lemma~\ref{lemma:convergence}), $q_t$ now converges to $d^*$, i.e., a profile between initial qualifications $q_0$ and ideal profile $d$, which is closer to $q_0$ with smaller investment $k$. It shows that if agents do not exert enough effort and the acquired knowledge is not sufficient, then they will not make satisfactory improvements.

\paragraph{Agent's utility and improvement action.} Denote agent utility under the forgetting mechanism as $\widehat{U}(k, \theta, r, x_0)$. Unlike settings without forgetting where we can derive the exact time $H$ it takes for agents to be accepted and find utility $U$ (\eqref{eq:utility}), the analytical form of $\widehat{U}(k, \theta, r, x_0)$ is not easy to derive. Nonetheless, we can still show that there exist scenarios under which agents have incentives to improve, 
even though the best attainable profile is a profile $d^*$ between initial $q_0$ and the ideal $d$. 
\begin{theorem}\label{theorem: fimprove}
For any threshold $\theta$ (resp. discounting factor $r$), there exists a discounting factor $r$ (resp. threshold $\theta$) such that agent's utility $\widehat{U}(\bar{k}, \theta, r, x_0) > 0$ for some $\bar{k}\in(0,\widehat{k})$, i.e., agents have the incentive to make a positive effort. The upper bound of the optimal effort is $\widehat{k}$ given by
\begin{align}\label{eq:khatdef}
\widehat{k} = \min\left(\frac{\widehat{x}_0^{2}}{2\widehat{x}_0^{2} + 2\widehat{x}_0^{3}}, \frac{x_0\cdot(x_0^2+x_0 - \sqrt{x_0^4-x_0^2+1})}{2x_0^2+2x_0^3-1}\right)
\end{align}
where $\widehat{x}_0$ is the root of $2x_0^2+2x_0^3-1 = 0$ within $(0,1)$.

\end{theorem}

Thm.~\ref{theorem: fimprove} implies that there exists $(\theta,r)$ such that agents best respond by improving their qualifications, and the optimal effort is upper bounded by $\widehat{k}$. 
Indeed, we can numerically find the upper bound $\widehat{k}$ as a function $x_0$ (shown in Fig.~\ref{fig:khat}). Because $\widehat{k}<0.35$ for all $x_0$,  the actual effort invested by any agent is less than 0.35, and the qualifications $q_t$ converge to a profile $d^*$ that is between $q_0$ and $0.35\cdot d+0.65\cdot q_0$. This means the improvement an agent can make under the forgetting mechanism may be limited, suggesting that the agents may not improve to be qualified when the tasks are challenging.

\section{Experiments}\label{sec:exp}

We validate theoretical results by conducting experiments on Exam score \citep{royce2012} and FICO score \citep{Fed2007} dataset\footnote{https://github.com/osu-srml/Alg-Persistent-Improvement}. For both datasets, scores serve as the agent's initial similarity $x_0$, and we assume agents interact with a decision maker based on the Stackelberg game in Sec.~\ref{sec:problem}. We first fit these scores with beta distributions, i.e., $x_0 \sim \text{Beta}(v,w)$, and then use them to derive the followings:
\begin{enumerate}[leftmargin=*,topsep=-0.em,itemsep=-0.em]
    \item The optimal decision threshold  $\theta^*$ for the decision-maker to incentivize the largest amount of improvement and promote \textit{social welfare}, and the total improvement induced by $\theta^*$.
    \item The percentage of agents who choose to manipulate under the decision-maker's optimal policy.
\end{enumerate}

\paragraph{Exam Score Data.} It is a synthetic dataset containing 1000 students' exam scores on 3 subjects including math, reading, and writing \citep{royce2012}. We first average over 3 subjects and normalize the averaged score to $[0,1]$. Then, we fit two beta distributions to the normalized scores of males and females and obtain $x_0\sim \text{Beta}(4.86,2.37), \text{Beta}(4.15,1.79)$ (see Fig.~\ref{fig:examdist} in App.~\ref{app:exp}). 

\begin{figure*}[h]
\centering
\includegraphics[trim=0.2cm 0.4cm 0.3cm 0.2cm,clip,width=0.248\textwidth]{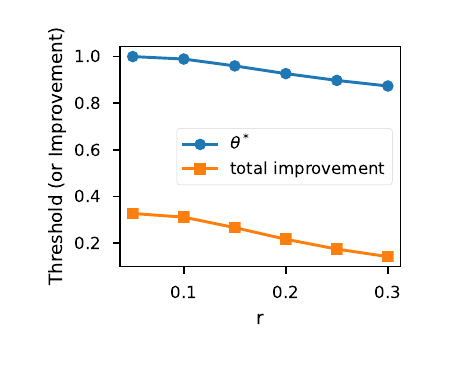}
\hspace{-0.2cm}
\includegraphics[trim=0.2cm 0.4cm 0.3cm 0.2cm,clip,width=0.248\textwidth]{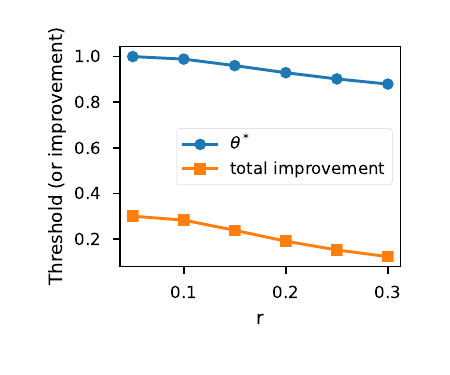}
\hspace{-0.2cm}
\includegraphics[trim=0.2cm 0.4cm 0.3cm 0.2cm,clip,width=0.248\textwidth]{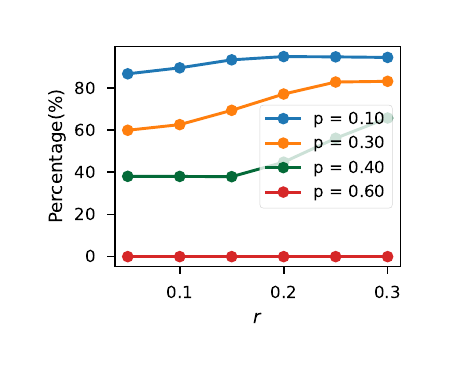}
\hspace{-0.2cm}
\includegraphics[trim=0.2cm 0.4cm 0.3cm 0.2cm,clip,width=0.248\textwidth]{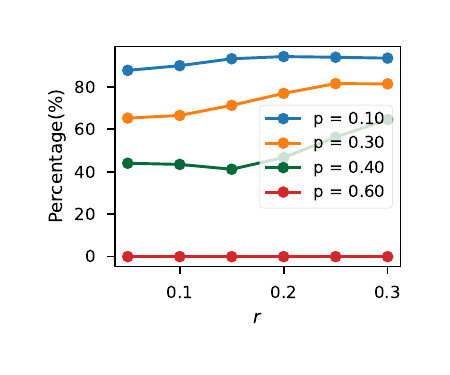}
\caption{From the left to the right are: optimal thresholds to incentivize improvement for males/females; manipulation probability under the thresholds for males/females for \textbf{Exam data}.}
\vspace{0.32cm}
\label{fig:optthreshold}
\end{figure*}

With these distributions, we can compute the optimal decision thresholds and the corresponding total improvement under different discounting factors $r$. As shown in  Fig.~\ref{fig:optthreshold}, for both males and females, the experimental results are similar. When $r$ increases, $\theta^*$ always decreases and the total amount of improvement becomes lower. This illustrates how larger discounting factors harm agents' improvement. Additionally, we consider settings with both manipulation and improvement. Fig.~\ref{fig:optthreshold} also shows the percentages of agents who prefer to manipulate under $\theta^*$. It shows that agents are less likely to manipulate as detection probability $P$ increases. 

\begin{figure*}[h]
\centering
\includegraphics[trim=0.2cm 0.2cm 0.4cm 0.2cm,clip,width=0.98\textwidth]{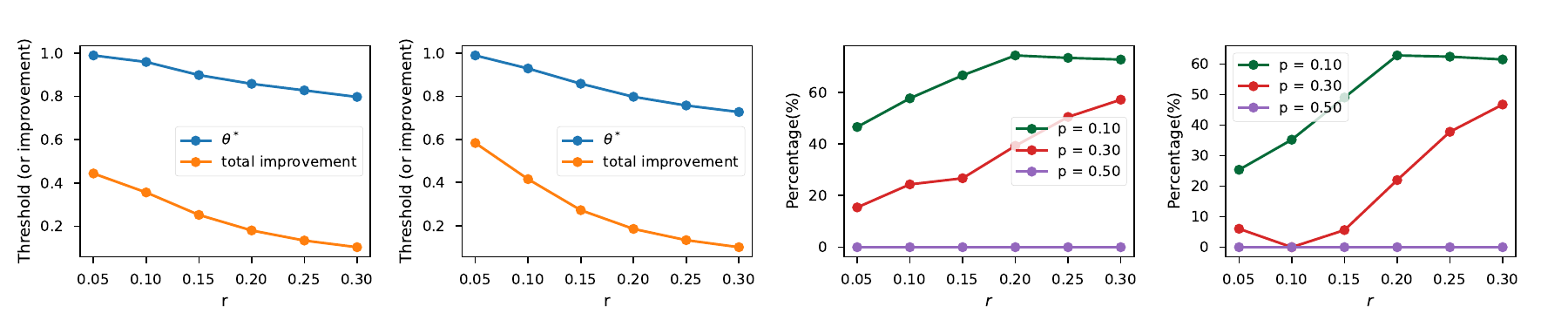}
\caption{From the left to the right are: optimal thresholds to incentivize improvement for Caucasians and African Americans; manipulation probability under the thresholds for Caucasians and African Americans for \textbf{FICO data}. }
\includegraphics[trim=0.2cm 0.2cm 0.4cm 0.2cm,clip,width=0.98\textwidth]{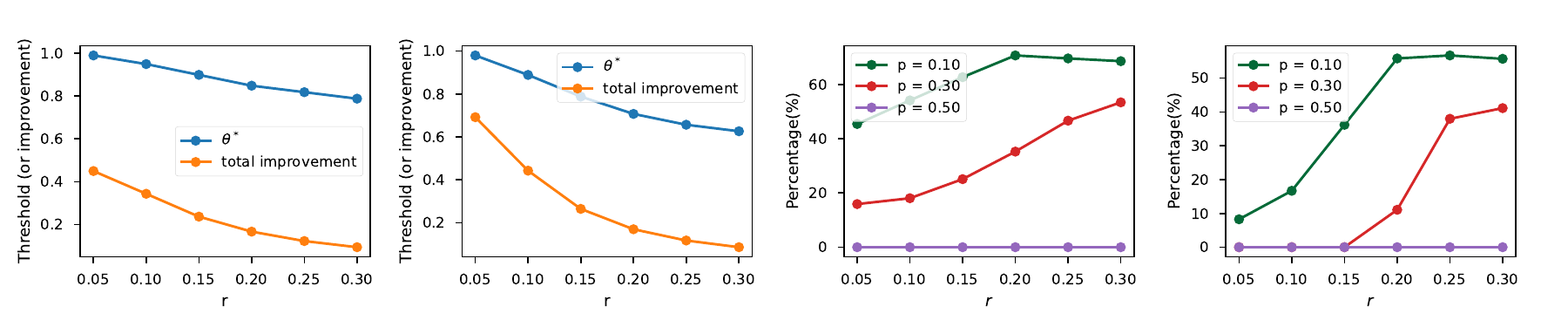}
\caption{Optimal thresholds to incentivize improvement (left two plots) and manipulation probability under the thresholds (right two plots) for Asians and Hispanic of the \textbf{FICO data}.}
\label{fig:optfico}
\end{figure*}

\paragraph{FICO Score Data.} We adopt the data pre-processed by \citet{Hardt2016b}, which contains CDF of credit scores of four racial groups (Caucasian, African American, Hispanic, Asian). For each group, we fit a Beta distribution and obtain four distributions: $\text{Beta}(1.11,0.97)$ for Caucasian, $\text{Beta}(0.91,3.84)$ for African American, $\text{Beta}(0.99,1.58)$ for Hispanic, $\text{Beta}(1.35,1.13)$ for Asian (see Fig.~\ref{fig:ficodist} in App. \ref{app:exp}). The results for Caucasians and African Americans are shown on the first column in Fig. \ref{fig:optfico}, while the results for Asian and Hispanic are shown the second column.

For each group, we compute the optimal decision threshold and corresponding total improvement under different $r$. As shown in Fig.~\ref{fig:optfico} (left two plots), for both groups, their corresponding optimal threshold $\theta^*$ and the total amount of improvement always decrease as $r$ increases. 
For settings with both manipulation and improvement (right two plots in Fig.~\ref{fig:optfico}), agents are more likely to manipulate under smaller detection probability. When detection probability $P$ is sufficiently large, agents do not have incentives to manipulate.

\section{Conclusions \& Limitations}\label{sec:limit}

This paper studies the strategic interactions between agents and a decision-maker when agent action has delayed and persistent effects. By utilizing a qualification dynamics model and the discounting utility function, we analyze the conditions where agents tend to improve and investigate how the decision-maker can incentivize agents to make the largest improvement. Moreover, we consider the situation where agents can improve or manipulate, and characterize how agents would make improvement or manipulation decisions when their efforts take time to pay back. Finally, we discuss the situation where the tasks are challenging and a forgetting mechanism takes place, thereby expanding the scope of our model.

However, our theoretical results depend on the assumption that both agents and the decision-maker have perfect information about each other so that they always best respond. Extension to cases when each party only has partial or imperfect information is important. Moreover, these theorems are based on the qualification dynamics~\eqref{eq:dynamics}. Although a scenario when it does not hold is studied in Sec.~\ref{sec:forget}, future works should also consider other variants tailored to specific applications to prevent negative outcomes. For example, they may consider the more complex situation where the decision-maker has a time-varying ideal profile $d_t$.

\vspace{-0.1cm}

\section*{Ethical Consideration Statement}

We believe our work fills the gap in which agent strategic behaviors are benign and agents' efforts can have long-lasting but diminishing effects. This can be the case under many real-world situations including exam preparation and job application. Thus, our model can improve trustworthy machine learning and decision-making in reality. However, as mentioned in Sec. \ref{sec:limit}, our work relies on certain assumptions and needs to be used cautiously. Moreover, though we provide a procedure to estimate the discounting factor, performing controlled experiments is not always accessible. Meanwhile, manipulation cost and detection probability are unknown and hard to estimate. Collecting real data and estimating these parameters remain promising research directions in the future.

\section*{Acknowledgement}

This material is based upon work supported by the U.S. National Science Foundation under award IIS-2202699, by grants from the Ohio State University's Translational Data Analytics Institute and College of Engineering Strategic Research Initiative.

\bibliography{aaai24}

\begin{thebibliography}{57}
\providecommand{\natexlab}[1]{#1}

\bibitem[{Ahmadi et~al.(2021)Ahmadi, Beyhaghi, Blum, and Naggita}]{ahmadi2021strategic}
Ahmadi, S.; Beyhaghi, H.; Blum, A.; and Naggita, K. 2021.
\newblock The strategic perceptron.
\newblock In \emph{Proceedings of the 22nd ACM Conference on Economics and Computation}, 6--25.

\bibitem[{Ahmadi et~al.(2022{\natexlab{a}})Ahmadi, Beyhaghi, Blum, and Naggita}]{ahmadi2022}
Ahmadi, S.; Beyhaghi, H.; Blum, A.; and Naggita, K. 2022{\natexlab{a}}.
\newblock On classification of strategic agents who can both game and improve.
\newblock \emph{arXiv preprint arXiv:2203.00124}.

\bibitem[{Ahmadi et~al.(2022{\natexlab{b}})Ahmadi, Beyhaghi, Blum, and Naggita}]{ahmadi2022setting}
Ahmadi, S.; Beyhaghi, H.; Blum, A.; and Naggita, K. 2022{\natexlab{b}}.
\newblock Setting Fair Incentives to Maximize Improvement.
\newblock \emph{arXiv preprint arXiv:2203.00134}.

\bibitem[{Alon et~al.(2020)Alon, Dobson, Procaccia, Talgam-Cohen, and Tucker-Foltz}]{Alon2020}
Alon, T.; Dobson, M.; Procaccia, A.; Talgam-Cohen, I.; and Tucker-Foltz, J. 2020.
\newblock Multiagent Evaluation Mechanisms.
\newblock \emph{Proceedings of the AAAI Conference on Artificial Intelligence}, 34: 1774--1781.

\bibitem[{Barsotti, Kocer, and Santos(2022)}]{Barsotti2022}
Barsotti, F.; Kocer, R.~G.; and Santos, F.~P. 2022.
\newblock Transparency, Detection and Imitation in Strategic Classification.
\newblock In \emph{Proceedings of the Thirty-First International Joint Conference on Artificial Intelligence, {IJCAI-22}}, 67--73.

\bibitem[{Bechavod et~al.(2021)Bechavod, Ligett, Wu, and Ziani}]{bechavod2021}
Bechavod, Y.; Ligett, K.; Wu, S.; and Ziani, J. 2021.
\newblock Gaming helps! learning from strategic interactions in natural dynamics.
\newblock In \emph{International Conference on Artificial Intelligence and Statistics}, 1234--1242.

\bibitem[{Bechavod et~al.(2022)Bechavod, Podimata, Wu, and Ziani}]{bechavod2022information}
Bechavod, Y.; Podimata, C.; Wu, S.; and Ziani, J. 2022.
\newblock Information discrepancy in strategic learning.
\newblock In \emph{International Conference on Machine Learning}, 1691--1715.

\bibitem[{Ben-Porat and Tennenholtz(2017)}]{Ben2017}
Ben-Porat, O.; and Tennenholtz, M. 2017.
\newblock Best Response Regression.
\newblock In \emph{Advances in Neural Information Processing Systems}.

\bibitem[{Braverman and Garg(2020)}]{Braverman2020}
Braverman, M.; and Garg, S. 2020.
\newblock The Role of Randomness and Noise in Strategic Classification.
\newblock \emph{CoRR}, abs/2005.08377.

\bibitem[{Castellano, Fortunato, and Loreto(2009)}]{castellano2009statistical}
Castellano, C.; Fortunato, S.; and Loreto, V. 2009.
\newblock Statistical physics of social dynamics.
\newblock \emph{Reviews of modern physics}, 81(2): 591.

\bibitem[{Chen, Wang, and Liu(2020)}]{Chen2020}
Chen, Y.; Wang, J.; and Liu, Y. 2020.
\newblock Strategic Recourse in Linear Classification.
\newblock \emph{CoRR}, abs/2011.00355.

\bibitem[{Dean et~al.(2024{\natexlab{a}})Dean, Curmei, Ratliff, Morgenstern, and Fazel}]{dean2024emergent}
Dean, S.; Curmei, M.; Ratliff, L.; Morgenstern, J.; and Fazel, M. 2024{\natexlab{a}}.
\newblock Emergent specialization from participation dynamics and multi-learner retraining.
\newblock In \emph{International Conference on Artificial Intelligence and Statistics}, 343--351. PMLR.

\bibitem[{Dean et~al.(2024{\natexlab{b}})Dean, Dong, Jagadeesan, and Leqi}]{dean2024accounting}
Dean, S.; Dong, E.; Jagadeesan, M.; and Leqi, L. 2024{\natexlab{b}}.
\newblock Accounting for AI and Users Shaping One Another: The Role of Mathematical Models.
\newblock \emph{arXiv preprint arXiv:2404.12366}.

\bibitem[{Dean and Morgenstern(2022)}]{Dean2022}
Dean, S.; and Morgenstern, J. 2022.
\newblock Preference Dynamics Under Personalized Recommendations.
\newblock In \emph{Proceedings of the 23rd ACM Conference on Economics and Computation}, 795–816.

\bibitem[{Dong et~al.(2018)Dong, Roth, Schutzman, Waggoner, and Wu}]{Dong2018}
Dong, J.; Roth, A.; Schutzman, Z.; Waggoner, B.; and Wu, Z.~S. 2018.
\newblock Strategic Classification from Revealed Preferences.
\newblock In \emph{Proceedings of the 2018 ACM Conference on Economics and Computation}, 55–70.

\bibitem[{Eilat et~al.(2022)Eilat, Finkelshtein, Baskin, and Rosenfeld}]{Eilat2022}
Eilat, I.; Finkelshtein, B.; Baskin, C.; and Rosenfeld, N. 2022.
\newblock Strategic Classification with Graph Neural Networks.

\bibitem[{Gaitonde, Kleinberg, and Tardos(2021)}]{gaitonde2021polarization}
Gaitonde, J.; Kleinberg, J.; and Tardos, {\'E}. 2021.
\newblock Polarization in geometric opinion dynamics.
\newblock In \emph{Proceedings of the 22nd ACM Conference on Economics and Computation}, 499--519.

\bibitem[{Gr{\"u}ne-Yanoff(2015)}]{grune2015models}
Gr{\"u}ne-Yanoff, T. 2015.
\newblock Models of temporal discounting 1937--2000: An interdisciplinary exchange between economics and psychology.
\newblock \emph{Science in context}, 28(4): 675--713.

\bibitem[{Hardt, Jagadeesan, and Mendler-D{\"u}nner(2022)}]{hardt2022p}
Hardt, M.; Jagadeesan, M.; and Mendler-D{\"u}nner, C. 2022.
\newblock Performative Power.
\newblock In \emph{Advances in Neural Information Processing Systems}.

\bibitem[{Hardt et~al.(2016{\natexlab{a}})Hardt, Megiddo, Papadimitriou, and Wootters}]{Hardt2016a}
Hardt, M.; Megiddo, N.; Papadimitriou, C.; and Wootters, M. 2016{\natexlab{a}}.
\newblock Strategic Classification.
\newblock In \emph{Proceedings of the 2016 ACM Conference on Innovations in Theoretical Computer Science}, 111–122.

\bibitem[{Hardt et~al.(2016{\natexlab{b}})Hardt, Price, Price, and Srebro}]{Hardt2016b}
Hardt, M.; Price, E.; Price, E.; and Srebro, N. 2016{\natexlab{b}}.
\newblock Equality of Opportunity in Supervised Learning.
\newblock In \emph{Advances in Neural Information Processing Systems}.

\bibitem[{Harris, Heidari, and Wu(2021)}]{harris2021stateful}
Harris, K.; Heidari, H.; and Wu, S.~Z. 2021.
\newblock Stateful strategic regression.
\newblock \emph{Advances in Neural Information Processing Systems}, 28728--28741.

\bibitem[{Hashimoto et~al.(2018)Hashimoto, Srivastava, Namkoong, and Liang}]{hashimoto2018fairness}
Hashimoto, T.; Srivastava, M.; Namkoong, H.; and Liang, P. 2018.
\newblock Fairness without demographics in repeated loss minimization.
\newblock In \emph{International Conference on Machine Learning}, 1929--1938. PMLR.

\bibitem[{Holmstrom and Milgrom(1991)}]{holmstrom1991multitask}
Holmstrom, B.; and Milgrom, P. 1991.
\newblock Multitask principal--agent analyses: Incentive contracts, asset ownership, and job design.
\newblock \emph{The Journal of Law, Economics, and Organization}, 24--52.

\bibitem[{Horowitz and Rosenfeld(2023)}]{horowitz2023causal}
Horowitz, G.; and Rosenfeld, N. 2023.
\newblock Causal Strategic Classification: A Tale of Two Shifts.
\newblock arXiv:2302.06280.

\bibitem[{Izzo, Ying, and Zou(2021)}]{Izzo2021}
Izzo, Z.; Ying, L.; and Zou, J. 2021.
\newblock How to Learn when Data Reacts to Your Model: Performative Gradient Descent.
\newblock In \emph{Proceedings of the 38th International Conference on Machine Learning}, 4641--4650.

\bibitem[{Jagadeesan, Mendler-D{\"u}nner, and Hardt(2021)}]{Hardt2021}
Jagadeesan, M.; Mendler-D{\"u}nner, C.; and Hardt, M. 2021.
\newblock Alternative Microfoundations for Strategic Classification.
\newblock In \emph{Proceedings of the 38th International Conference on Machine Learning}, 4687--4697.

\bibitem[{Jin et~al.(2024{\natexlab{a}})Jin, Xie, Liu, and Zhang}]{jin2024addressing}
Jin, K.; Xie, T.; Liu, Y.; and Zhang, X. 2024{\natexlab{a}}.
\newblock Addressing Polarization and Unfairness in Performative Prediction.
\newblock \emph{arXiv preprint arXiv:2406.16756}.

\bibitem[{Jin et~al.(2024{\natexlab{b}})Jin, Yin, Chen, Sun, Zhang, Liu, and Liu}]{jin2024performative}
Jin, K.; Yin, T.; Chen, Z.; Sun, Z.; Zhang, X.; Liu, Y.; and Liu, M. 2024{\natexlab{b}}.
\newblock Performative federated learning: A solution to model-dependent and heterogeneous distribution shifts.
\newblock In \emph{Proceedings of the AAAI Conference on Artificial Intelligence}, volume~38, 12938--12946.

\bibitem[{Jin et~al.(2022)Jin, Zhang, Khalili, Naghizadeh, and Liu}]{Jin2022}
Jin, K.; Zhang, X.; Khalili, M.~M.; Naghizadeh, P.; and Liu, M. 2022.
\newblock Incentive Mechanisms for Strategic Classification and Regression Problems.
\newblock In \emph{Proceedings of the 23rd ACM Conference on Economics and Computation}, 760–790.

\bibitem[{Kaelbling, Littman, and Moore(1996)}]{kaelbling1996}
Kaelbling, L.~P.; Littman, M.~L.; and Moore, A.~W. 1996.
\newblock Reinforcement learning: A survey.
\newblock \emph{Journal of artificial intelligence research}, 4: 237--285.

\bibitem[{Kimmons(2012)}]{royce2012}
Kimmons, R. 2012.
\newblock Synthetic Exam Scores in a Public School.
\newblock Technical report, Brigham Young University.

\bibitem[{Kleinberg and Raghavan(2020)}]{Kleinberg2020}
Kleinberg, J.; and Raghavan, M. 2020.
\newblock How Do Classifiers Induce Agents to Invest Effort Strategically?
\newblock 1–23.

\bibitem[{Krahn and Gafni(1993)}]{krahn1993}
Krahn, M.; and Gafni, A. 1993.
\newblock Discounting in the economic evaluation of health care interventions.
\newblock \emph{Medical care}, 403--418.

\bibitem[{Lechner, Urner, and Ben-David(2023)}]{lechner2023strategic}
Lechner, T.; Urner, R.; and Ben-David, S. 2023.
\newblock Strategic classification with unknown user manipulations.
\newblock In \emph{International Conference on Machine Learning}, 18714--18732. PMLR.

\bibitem[{Liu et~al.(2019)Liu, Dean, Rolf, Simchowitz, and Hardt}]{Lydia2019}
Liu, L.~T.; Dean, S.; Rolf, E.; Simchowitz, M.; and Hardt, M. 2019.
\newblock Delayed Impact of Fair Machine Learning.
\newblock In \emph{Proceedings of the Twenty-Eighth International Joint Conference on Artificial Intelligence, {IJCAI-19}}, 6196--6200.

\bibitem[{Liu, Garg, and Borgs(2022)}]{liu22}
Liu, L.~T.; Garg, N.; and Borgs, C. 2022.
\newblock Strategic ranking.
\newblock In \emph{Proceedings of The 25th International Conference on Artificial Intelligence and Statistics}, 2489--2518.

\bibitem[{Meier and Sprenger(2013)}]{meier2013}
Meier, S.; and Sprenger, C.~D. 2013.
\newblock Discounting financial literacy: Time preferences and participation in financial education programs.
\newblock \emph{Journal of Economic Behavior \& Organization}, 95: 159--174.

\bibitem[{Miehling et~al.(2019)Miehling, Dong, Langbort, and Ba{\c{s}}ar}]{miehling2019strategic}
Miehling, E.; Dong, R.; Langbort, C.; and Ba{\c{s}}ar, T. 2019.
\newblock Strategic inference with a single private sample.
\newblock In \emph{2019 IEEE 58th Conference on Decision and Control (CDC)}, 2188--2193. IEEE.

\bibitem[{Miller, Milli, and Hardt(2020)}]{Miller2020}
Miller, J.; Milli, S.; and Hardt, M. 2020.
\newblock Strategic Classification is Causal Modeling in Disguise.
\newblock In \emph{Proceedings of the 37th International Conference on Machine Learning}.

\bibitem[{Perdomo et~al.(2020)Perdomo, Zrnic, Mendler-D{\"u}nner, and Hardt}]{perdomo2020}
Perdomo, J.; Zrnic, T.; Mendler-D{\"u}nner, C.; and Hardt, M. 2020.
\newblock Performative Prediction.
\newblock In \emph{Proceedings of the 37th International Conference on Machine Learning}, 7599--7609.

\bibitem[{Raab and Liu(2021)}]{raab2021unintended}
Raab, R.; and Liu, Y. 2021.
\newblock Unintended selection: Persistent qualification rate disparities and interventions.
\newblock \emph{Advances in Neural Information Processing Systems}, 26053--26065.

\bibitem[{Reserve(2007)}]{Fed2007}
Reserve, U.~F. 2007.
\newblock Report to the congress on credit scoring and its effects on the availability and affordability of credit.
\newblock In \emph{Board of Governors of the Federal Reserve System}.

\bibitem[{Rosenfeld et~al.(2020)Rosenfeld, Hilgard, Ravindranath, and Parkes}]{Rose2020}
Rosenfeld, N.; Hilgard, A.; Ravindranath, S.~S.; and Parkes, D.~C. 2020.
\newblock From Predictions to Decisions: Using Lookahead Regularization.
\newblock In \emph{Advances in Neural Information Processing Systems}, 4115--4126.

\bibitem[{Samuelson(1937)}]{samuelson1937note}
Samuelson, P.~A. 1937.
\newblock A note on measurement of utility.
\newblock \emph{The review of economic studies}, 4(2): 155--161.

\bibitem[{Shao, Blum, and Montasser(2024)}]{shao2024strategic}
Shao, H.; Blum, A.; and Montasser, O. 2024.
\newblock Strategic classification under unknown personalized manipulation.
\newblock \emph{Advances in Neural Information Processing Systems}, 36.

\bibitem[{Shavit, Edelman, and Axelrod(2020)}]{shavit2020}
Shavit, Y.; Edelman, B.~L.; and Axelrod, B. 2020.
\newblock Causal Strategic Linear Regression.
\newblock In \emph{Proceedings of the 37th International Conference on Machine Learning}, ICML'20.

\bibitem[{Sundaram et~al.(2021)Sundaram, Vullikanti, Xu, and Yao}]{Sundaram2021}
Sundaram, R.; Vullikanti, A.; Xu, H.; and Yao, F. 2021.
\newblock PAC-Learning for Strategic Classification.
\newblock In \emph{Proceedings of the 38th International Conference on Machine Learning}, 9978--9988.

\bibitem[{Xie and Zhang(2024{\natexlab{a}})}]{xie2024automating}
Xie, T.; and Zhang, X. 2024{\natexlab{a}}.
\newblock Automating Data Annotation under Strategic Human Agents: Risks and Potential Solutions.
\newblock \emph{arXiv preprint arXiv:2405.08027}.

\bibitem[{Xie and Zhang(2024{\natexlab{b}})}]{xie2024non}
Xie, T.; and Zhang, X. 2024{\natexlab{b}}.
\newblock Non-linear Welfare-Aware Strategic Learning.
\newblock \emph{arXiv preprint arXiv:2405.01810}.

\bibitem[{Xie et~al.(2024)Xie, Zuo, Khalili, and Zhang}]{xie2024learning}
Xie, T.; Zuo, Z.; Khalili, M.~M.; and Zhang, X. 2024.
\newblock Learning under Imitative Strategic Behavior with Unforeseeable Outcomes.
\newblock \emph{arXiv preprint arXiv:2405.01797}.

\bibitem[{Yin et~al.(2024)Yin, Raab, Liu, and Liu}]{yin2024long}
Yin, T.; Raab, R.; Liu, M.; and Liu, Y. 2024.
\newblock Long-term fairness with unknown dynamics.
\newblock \emph{Advances in Neural Information Processing Systems}, 36.

\bibitem[{Zhang et~al.(2022)Zhang, Khalili, Jin, Naghizadeh, and Liu}]{zhang2022}
Zhang, X.; Khalili, M.~M.; Jin, K.; Naghizadeh, P.; and Liu, M. 2022.
\newblock Fairness interventions as (dis) incentives for strategic manipulation.
\newblock In \emph{International Conference on Machine Learning}, 26239--26264. PMLR.

\bibitem[{Zhang, Khalili, and Liu(2020)}]{zhang2020long}
Zhang, X.; Khalili, M.~M.; and Liu, M. 2020.
\newblock Long-term impacts of fair machine learning.
\newblock \emph{ergonomics in design}, 28(3): 7--11.

\bibitem[{Zhang et~al.(2019)Zhang, Khalili, Tekin, and Liu}]{zhang2019group}
Zhang, X.; Khalili, M.~M.; Tekin, C.; and Liu, M. 2019.
\newblock Group retention when using machine learning in sequential decision making: the interplay between user dynamics and fairness.
\newblock In \emph{Proceedings of the 33rd International Conference on Neural Information Processing Systems}, 15269--15278.

\bibitem[{Zhang and Liu(2021)}]{zhang2021fairness}
Zhang, X.; and Liu, M. 2021.
\newblock Fairness in learning-based sequential decision algorithms: A survey.
\newblock In \emph{Handbook of Reinforcement Learning and Control}, 525--555.

\bibitem[{Zhang et~al.(2020)Zhang, Tu, Liu, Liu, Kjellstrom, Zhang, and Zhang}]{zhang2020}
Zhang, X.; Tu, R.; Liu, Y.; Liu, M.; Kjellstrom, H.; Zhang, K.; and Zhang, C. 2020.
\newblock How do fair decisions fare in long-term qualification?
\newblock In \emph{Advances in Neural Information Processing Systems}, 18457--18469.

\end{thebibliography}

\appendix
\newpage
\onecolumn 

\section{Discussion and Generalization of \eqref{eq:dynamics}}\label{app:relax}

\paragraph{More details on the dynamics in \eqref{eq:dynamics}.} In the main paper, we assume the influence of the initial effort $k$ is persistent and will enable $q_t$ changes gradually during each round. This is well-supported by the following examples:

\begin{enumerate}
[leftmargin=*,topsep=-0.2em,itemsep=-0.2em]
\item \textit{Creditworthiness}: To improve creditworthiness, an individual may learn that an ideal profile would be a person with a constant high income and long-lasting good credit history. Therefore, she may exert a significant effort to find a job with a high salary. However, the effort will take several months or even one year for her to finally build up the ideal profile because she needs to work for a while to receive money and build a competitive credit history. 

\item \textit{Job application}: An individual who wants to apply for a technology company may learn about the skill set of an ideal candidate from several resources (e.g., the job description, alumni who work at the company, info session) and then exert a significant effort to study the required knowledge. However, it still takes time for her to do exercises and master the skills, resulting in a delay of finally being qualified. 
\end{enumerate}

\paragraph{Model generalization when agents can invest efforts at different time steps.} We discuss how the model in the main paper can capture more complicated scenarios where agents repeatedly exert efforts multiple times until they reach the target. Each effort has persistent effects on improving the qualification as shown in Eqn. \eqref{eq:complex}.
\begin{align}\label{eq:complex}
    \widetilde q_{t+1} = q_t + \sum_{\tau=0}^t k_\tau\cdot q_t^T d\cdot d \\
    \nonumber q_{t+1} = \frac{\widetilde q_{t+1}}{\| \widetilde q_{t+1} \|_2}
\end{align}
where $\sum_{\tau=0}^t k_\tau\in[0,1]$. This means the agents are able to invest more effort at arbitrary time steps (e.g., studying more skills in the middle of the preparing process), but the cumulative effort should not exceed 1 (they cannot master 110\% of knowledge). 

We first prove that there exists an effort $k^*\in[0,1]$ such that investing $k^{*}$ once at the beginning has the same impact on $\lim_{t\to \infty}q_t$ as investing a sequence of efforts $\{k_t\}_{t\geq0}$ over time: define $q^{min}_{t}, q^{max}_{t}$ as the "what-if" qualifications if the agents invest $k = 0$ or $k = 1$ at the initial round. Since $\sum_{\tau=0}^t k_\tau\in[0,1]$, we know the $q_t$ must be between $q_{min}^{t}, q_{max}^{t}$. Then because $q_t$ is continuous with respect to $k$, so we know $k^{*}$ must exist. Therefore, our model in the main paper can indeed assimilate the more complex setting.

\paragraph{Model generalization when $k$ diminishes with $t$.} In the main paper, $k_t$ is always equal to $k$, demonstrating the effort has a consistent and persistent effect on the improvement of an individual. According to Lemma \ref{lemma:convergence}, the similarity $x_t$ approaches $1$ at an exponential rate. Thus, the case of $k_t > k$ is not interesting since the convergence is faster and it may not make sense in practice that the effort can be increasingly effective as time goes on. However, in reality, it may be possible that $k_t$ is decreasing. This is a ``middle-point" case between the regular improvement in \eqref{eq:dynamics} and the forgetting mechanism \eqref{eq:fdynamic}, which may illustrate the ``tiredness" when agents stick to improve. However, we can prove that when $k_t$ decreases linearly (i.e., $k_t = \Theta(\frac{k}{t})$), the similarity $x_t$ can only converge to $1$ at a speed $\Theta(t^k)$.

\begin{theorem}\label{theorem:generalized}
    When $k_t$ decreases linearly (i.e., $k_t = \Theta(\frac{k}{t+1})$), $x_t$ converges to $1$ at a rate $\Theta(t^k)$
\end{theorem}

We prove Thm. \ref{theorem:generalized} in App. \ref{subsection:proofg}. Basically, this result illustrates that the agents will still improve to be qualified if $k_t$ decreases at a linear rate. Specifically, we can rewrite the \eqref{eq:convergence} as:

\begin{eqnarray}\label{eq:generalconvergence}
x_{t}^{-2} - 1 = \frac{(x_0)^{-2}-1}{(t+1)^{2k}}
\end{eqnarray}

From \eqref{eq:generalconvergence}, we can derive similar results of the agents' best responses and work out the thresholds for them to improve.

\section{Illustration of Table \ref{table: domain}}\label{app:corollary3}

\begin{figure}[h]\label{fig: illuc2}
     \centering
 \begin{subfigure}[b]{0.45\columnwidth}
     \centering        \includegraphics[width=\linewidth]{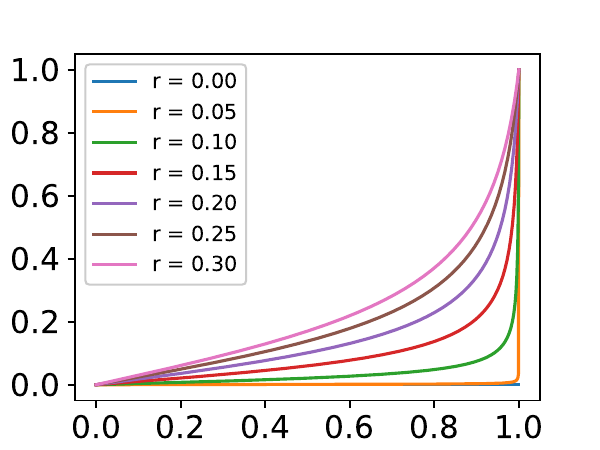}
         \caption{$x_0$ to improve under $(\theta, r)$}
     \end{subfigure}
     \hspace{0.5cm}
     \begin{subfigure}[b]{0.45\columnwidth}
     \centering       \includegraphics[width=\linewidth]{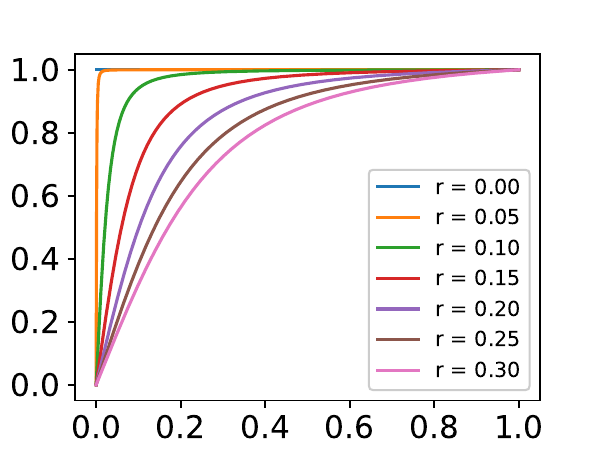}
         \caption{Best profile to reach for $x_0$}
     \end{subfigure}
     \caption{Illustration of Table \ref{table: domain}}\label{fig:corollary3}
     \centering
\end{figure}

Table \ref{table: domain} illustrate the minimum requirement of $x_0$ for an individual to improve under different $(\theta,r)$, and the best attainable profile for individuals with initial similarity  $x_0$. We illustrate them in Fig.~\ref{fig:corollary3}.

\paragraph{Discussions of intervention strategies in real applications.} Table \ref{table: domain} further suggest effective  strategies that encourage individuals to improve their qualifications, i.e., more individuals are incentivized to improve if (i) the decision-maker's acceptance threshold $\theta$ is lower; or (ii) the time it takes for individuals to succeed after investments is shorter. Examples of both strategies in real applications are as follows.
\begin{enumerate}[leftmargin=*,topsep=-0.2em,itemsep=-0.3em]
\item \textit{Lower acceptance threshold $\theta$ in hiring:} Instead of directly recruiting the qualified candidates,   companies first lower the standard by offering internship opportunities to encourage applicants to improve, and then offer full-time positions. This two-stage hiring process widens the candidate pool and incentivizes more people to improve.

 \item \textit{Lower discounting factor $r$ in college admission:} Instead of directly rejecting the unqualified high school graduates,  universities incentivize them by issuing {conditional transfer offers}. Once these students meet certain requirements, they get admitted. The conditional acceptances encourage more students to improve by lowering the time it takes for them to receive reward.
 
\end{enumerate}

Meanwhile, Table \ref{table: domain} also reveals that setting short-term goals will be effective to incentivize individuals to improve. For instance, teachers may set up several quizzes to break down the grade and make students more motivated to improve.

\section{Illustration of Thm.~\ref{theorem: choice}}\label{app:thm5}

\begin{figure}[H]
\centering
\includegraphics[trim=0.cm 0.6cm 0.2cm 0.9cm,clip,width=0.35\textwidth]{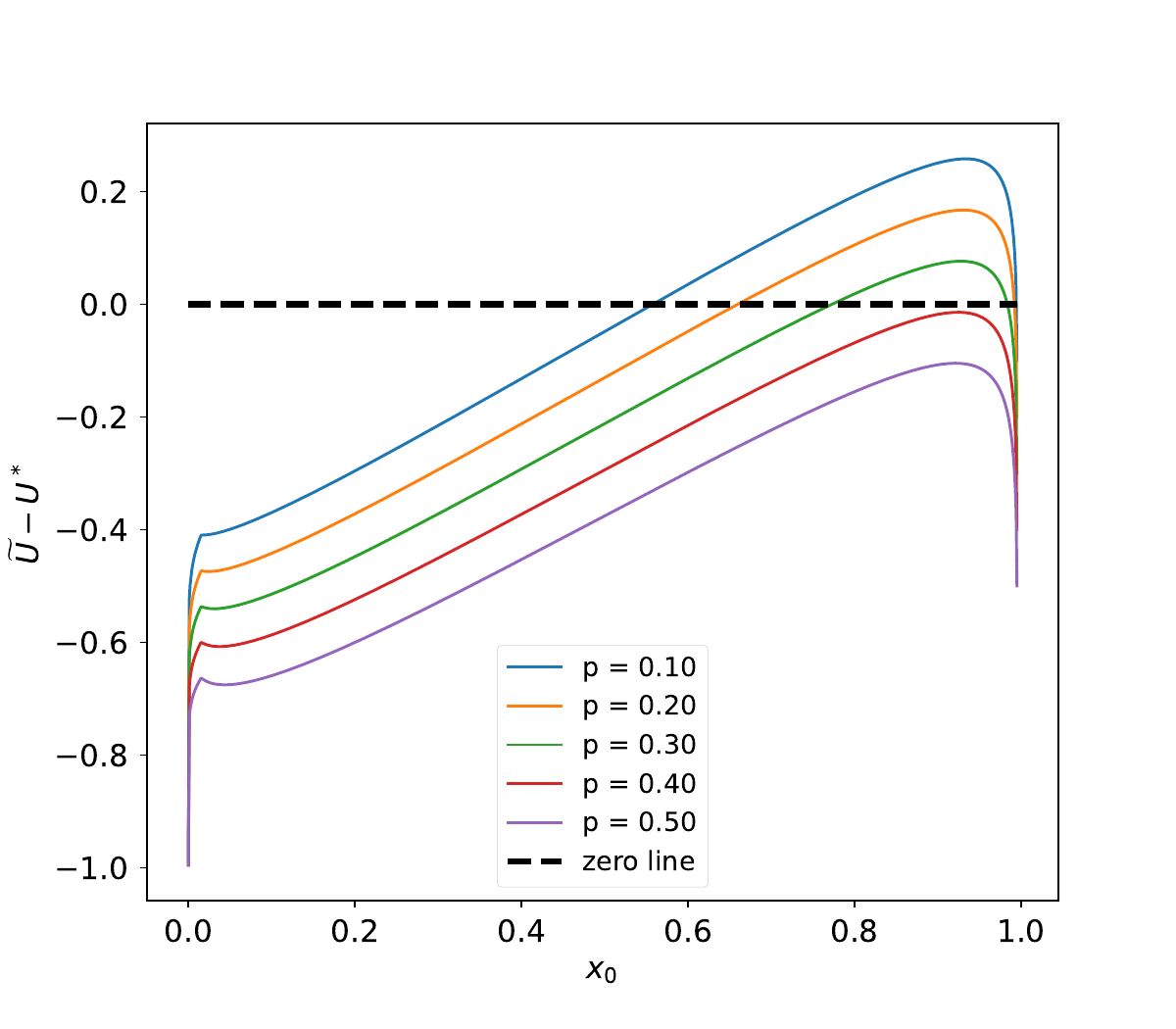}
\includegraphics[width=0.45\textwidth]{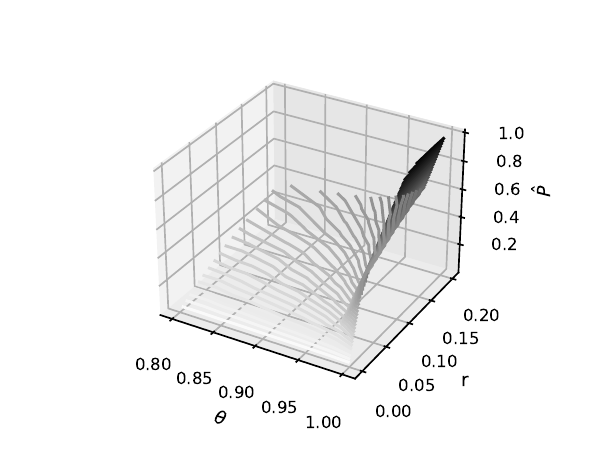}
\caption{Illustration of Thm. \ref{theorem: choice}: the left figure shows $\widetilde{U} - U$ as functions of $x_0$ under different $P$ when $\theta=0.995,r=0.05$;  the right plot shows threshold $\widehat{P}$ under different pairs of $(\theta, r)$. }
\label{fig:illu5}
\end{figure}

\begin{table*}[h]
\small
\begin{center}
\caption{Ranges $(\widehat{x_1},\widehat{x_2})$ of initial similarity $x_0$ under which individuals prefer to manipulate.}
\label{table: manip vs improve}
\resizebox{0.9\textwidth}{!}{
\begin{tabular}{cccccccc}\toprule
\multicolumn{1}{c}{\multirow{2}{*}{$\theta$}} & \multicolumn{1}{c}{\multirow{2}{*}{$r$}} 
 & \multicolumn{6}{c}{Detection probability $P$}  \\
  \cmidrule(rl){3-8} 
& &\makebox[3em]{0}&\makebox[3em]{0.1}&\makebox[3em]{0.2}
&\makebox[3em]{0.3}&\makebox[3em]{0.4}&\makebox[3em]{0.5}\\\midrule
$0.995$&$0.1$ &$(0.364,0.995)$&$(0.435,0.994)$&$(0.513,0.993)$&$(0.596,0.991)$&$(0.686,0.984)$&$(0.796,0.966)$\\ 
$0.976$&$0.05$ &$(0.499,0.976)$&$(0.613,0.973)$&$(0.740,0.958)$&$\emptyset$&$\emptyset$&$\emptyset$\\ 
$0.953$&$0.01$ &$(0.773,0.953)$&$\emptyset$&$\emptyset$&$\emptyset$&$\emptyset$&$\emptyset$\\\bottomrule
\end{tabular}}
\end{center}
\end{table*}
\vspace{-0.4cm}

Thm.~\ref{theorem: choice} identifies conditions under which manipulation (or improvement) is preferred by individuals over the other. As mentioned in Section~\ref{sec:manipulate}, the specific values of  $\widehat{P}$, $\widehat{x}$, $\widehat{x_1}$, $\widehat{x_2}$ in Thm.~\ref{theorem: choice} depend on $\theta$, $r$, and we can empirically find  $\widehat{P}$, $\widehat{x}$, $\widehat{x_1}$, $\widehat{x_2}$ and verify the theorem, as illustrated in Figure~\ref{fig:illu5} and Table~\ref{table: manip vs improve}. 
Specifically, the left plot in Figure~\ref{fig:illu5} shows $\widetilde{U} - U$ as functions of initial similarity $x_0$ under different detection probability $P$. Because individuals only prefer to manipulate if  $\widetilde{U} - U>0$, the plot shows the values of $\widehat{P}$, $\widehat{x}$, $\widehat{x_1}$, $\widehat{x_2}$ in Thm.~\ref{theorem: choice}. The right plot shows threshold $\widehat{P}$ under different pairs of $(\theta, r)$, and it shows that $\widehat{P}$ increases as $r$ increases. Table~\ref{table: manip vs improve} shows ranges $(\widehat{x_1},\widehat{x_2})$ of initial similarity $x_0$ under different detection probability $P$, acceptance threshold $\theta$, and discounting factor $r$.

\section{Additional Experiments}\label{app:exp}

\paragraph{Exam Score Data}

\begin{figure}
\centering
\includegraphics[width=0.3\textwidth]{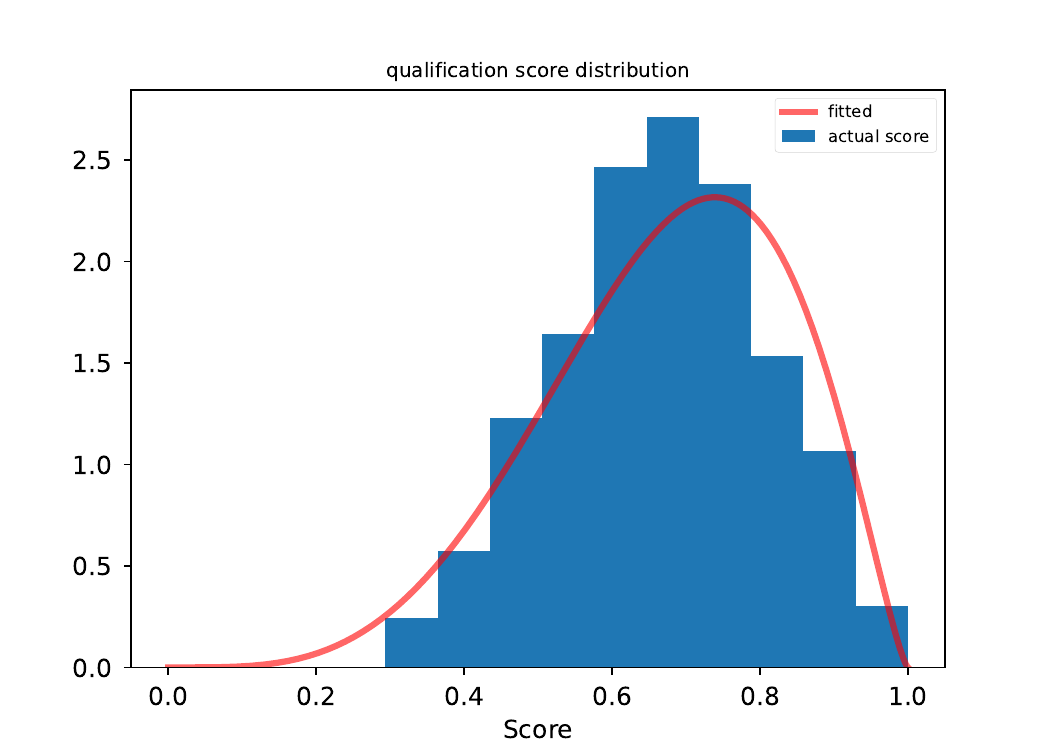}
\includegraphics[width=0.3\textwidth]{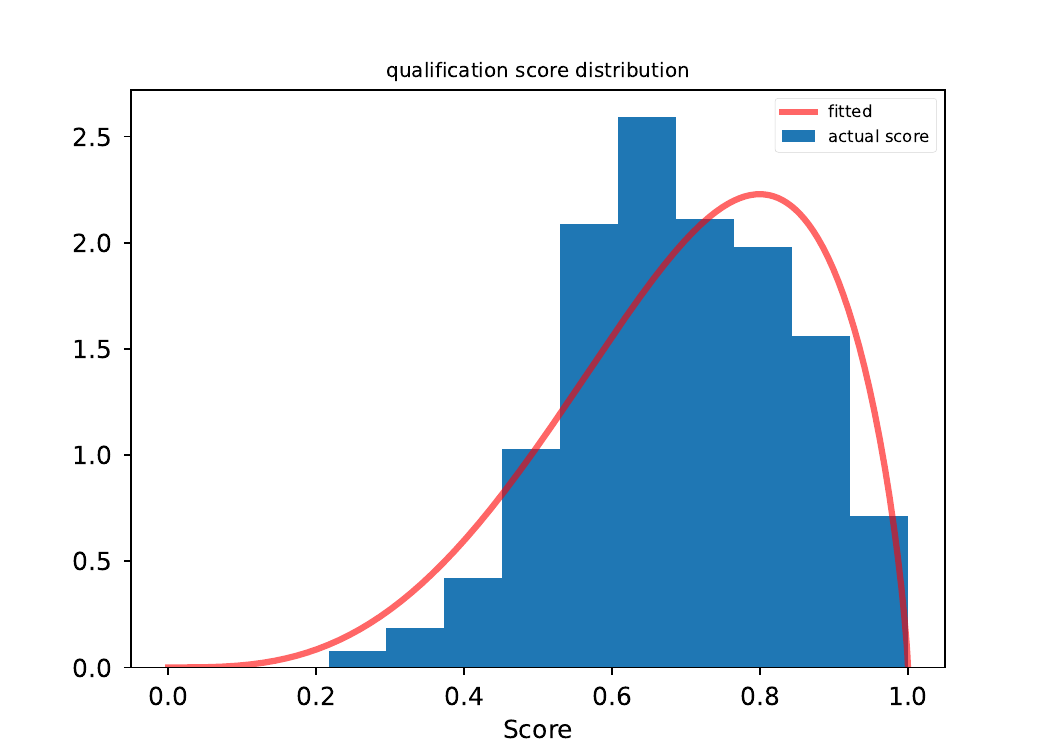}
\caption{Exam Score: $Beta$ distributions}
\label{fig:examdist}
\end{figure}

Just as Sec. \ref{sec:exp} mentions, we acquire the exam score data \cite{royce2012}, preprocess the data and fit beta distributions for both males and females. The fitted distribution and real distribution are shown in Fig. \ref{fig:examdist}.

\paragraph{FICO Score Data}

\begin{figure*}
\includegraphics[width=0.98\textwidth]{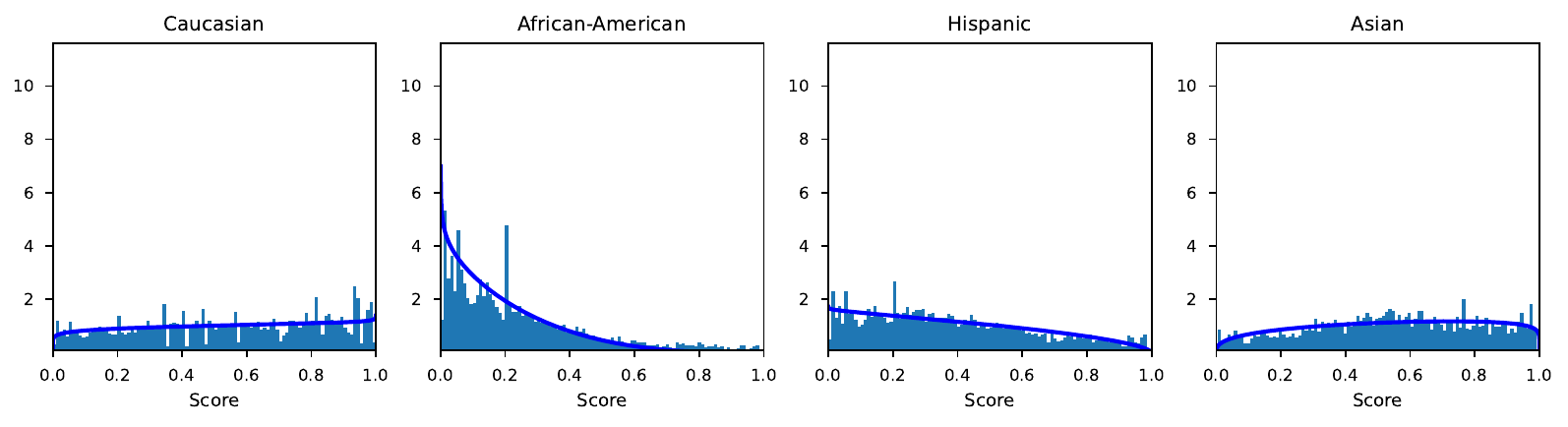}
\caption{FICO Score: Caucasian ($\text{Beta}(1.11,0.97)$), African American ($\text{Beta}(0.91,3.84)$), Hispanic ($\text{Beta}(0.99,1.58)$), Asian ($\text{Beta}(1.35,1.13)$)}
\label{fig:ficodist}
\end{figure*}

Just as Sec. \ref{sec:exp} mentions, we fit beta distributions for FICO Score \cite{Hardt2016b}, and obtain four distributions for different racial groups as shown in Fig. \ref{fig:ficodist}.

\section{Estimating the discounting factor $r$ in Sec.\ref{sec:educator}}\label{app:restimate}

We can estimate the discounting factor $r$ if given an experimental population. The decision-maker can publish an arbitrary threshold $\theta$ and observe the lowest score among all individuals who change their scores, which is $x^*(\theta)$. Then the decision-maker can use any expression in Table \ref{table: domain} to estimate $r$. Multiple experiments can make the estimation more robust.

\FloatBarrier

\section{Proofs}

\subsection{Proof Details of Thm. \ref{theorem: effort}}\label{subsection: proof1}


To derive $k^*$, we first take the derivative of \eqref{eq:utility} with respect to $k$. For simplicity, let $K = k+1$ and the derivative will not change. Also, let $R = r+1$ and $G = -\ln\left(\sqrt{\frac{(\theta)^{-2}-1}{(x_0)^{-2}-1}}\right)$. Then show the results as follows:

\abovedisplayskip=0pt
\belowdisplayskip=5pt

\begin{eqnarray}\label{eq: firstduhat}
\frac{\partial U}{\partial K} = \ln R \cdot R^{\frac{-G}{\ln K}} \cdot \frac{G}{K\cdot\ln^2K} - 1
\end{eqnarray}

\begin{eqnarray}\label{eq: secondduhat}
\frac{\partial^2 U}{\partial K^2} = \frac{-G\cdot\ln R \cdot R^{\frac{-G}{\ln K}}(\ln^2K+2\ln K -G \cdot \ln R)}{K^2\cdot\ln^4 K}
\end{eqnarray}

The denominator of $\frac{\partial^2 U}{\partial K^2}$ is always positive, and the first term $-G \cdot \ln R \cdot R^{\frac{-G}{\ln K}}$ of numerator is always negative. 

Also, because $K \in [1,2]$, $\ln^2K + 2\ln K\in(0,\ln^22 + 2\ln 2)$. Thus, we have following situations:
 
1) If $G \cdot \ln R > \ln^22+2\ln 2$, $\frac{\partial^2 U}{\partial K^2}$ is always positive when $K \in [1,2]$. This means $\frac{\partial U}{\partial K}$ is increasing. 

Then, noticing that $\lim_{K\to\ 1^+}\frac{\partial U}{\partial K} = -1$, we know $\frac{\partial U}{\partial K}$ is always negative when $K \in [1,2]$. This means $U$ is monotonically decreasing. Also, when $k = 0$, $U = 0$. This ensures $U$ is always non-positive and individuals will never choose to invest any effort.

2) If $G \cdot \ln R \le \ln^22+2\ln 2$, $\frac{\partial^2 U}{\partial K^2}$ is first positive, then negative when $K \in [1,2]$. Also, if plugging $K = 2$ into \eqref{eq: firstduhat}, we know $\lim_{K\to\ 2}\frac{\partial U}{\partial K} < 0$. These facts reveal that $\frac{\partial U}{\partial K}$ is firstly increasing from a negative number and then decreasing to a negative number. And there must exist a unique maximum point when $K = K^{'}$, $K^{'}$ should satisfy:

\begin{eqnarray}\label{eq:Kcondition}
\ln^2K^{'} + 2\ln K^{'} -G\cdot\ln R = 0
\end{eqnarray}

Plug \eqref{eq:Kcondition} into \eqref{eq: firstduhat}. Denote $\ln K^{'}$ as $t \in [0, \ln 2]$, and denote $\frac{\partial U}{\partial K}$ at $K^{'}$ as $L$:

\begin{eqnarray}\label{eq:tcondition0}
L = \frac{t+2}{t \cdot e^{2t+2}} - 1
\end{eqnarray}

Then take the derivative of $L$:

\begin{eqnarray}\label{eq:tcondition}
\frac{\partial L}{\partial t} = \frac{-2(t+1)^2\cdot e^{2t+2}}{t^2 \cdot e^{4t+4}} < 0
\end{eqnarray}

\eqref{eq:tcondition} shows $L$ is decreasing. Also, noticing that $\lim_{t\to\ 0^+}L(t) = +\infty$
and $\lim_{t\to\ \ln 2}L(t) < \frac{3}{2e^2} - 1 < 0$, we know there must exist a $t^{'} \in (0, \ln 2)$ as the root of $M$. We can explicitly solve $t^{'} = 0.1997$.

Thus, we now know that when $t \in [0, t^{'}]$, $L \ge 0$. With the plausible domain of $t$ and \eqref{eq:Kcondition}, we would know: When $G\ln R \in [0, t^{'2} + 2t^{'}]$, $L \ge 0$ and thereby $U$ has an extreme large point with value $U^*$. At this maximum point, \eqref{eq: firstduhat} equals 0, and \eqref{eq: secondduhat} is smaller than 0.

Finally, we derive the condition for $U^* > 0$: Denote $G\ln R$ as $C$ and $\ln K$ as $z$, $U$ can be simplified to:

\begin{eqnarray}\label{eq:usimple}
U = e^{\frac{-C}{z}} - e^z + 1
\end{eqnarray}

Because $z \in [0,\ln 2]$, for any $t$ fixed, $\lim_{C\to\ 0}U = 2 - e^z \ge 0$ and $\lim_{C\to\ 0}U = 1 - e^z \le 0$. With the fact that $\frac{\partial U}{\partial C} < 0$, we know $U$ is monotonically decreasing with $C$, so is $U^*$. Thus, there must exist a threshold $m$, when $C < m$, $U^* > 0$. And if $U^* > 0$, individuals will decide to improve. Then Thm. \ref{theorem: effort} is proved and we can numerically solve the threshold $m = 0.316$.

Although we believe exponential discounting is general and fits our setting well, we also note that we can still use derivative analysis when the discounting changes (e.g., hyperbolic discounting). Specifically, if denoting the discounted reward as $d(r,t)$, we would have $U = d(r,H) - k$. Then if taking the derivative we will get $\frac{\partial U}{\partial k} = \frac{\partial d}{\partial H}\cdot \frac{\partial H}{\partial k} - 1$. Noticing that $H$ is known, then discussing the properties of $d$ with different choices of discounting is enough to derive the nature of $U$. 

\subsection{Proof Details of Thm. \ref{theorem: optimal theta} and Corollary \ref{corollary: theta change vs r}}\label{subsection:proof1.5}

\subsubsection{Proof of Thm. \ref{theorem: optimal theta}}

First prove $U_d(\theta)$ has a maximize $\theta^* \in (0,1)$:

With the definition of $U_d(\theta)$ in \eqref{eq: Ud}, we already know $U_d$ is continuous. We can first observe that $U_d(0) = 0, U_d(1) = 1$. These hold simply because $x^*(0) = 0 and x^*(1) = 1$. Next noticing that for any $\theta \in (0,1)$, $U_d(\theta) > 0$ holds. This suggests that $\theta$ will reach its maximum point according to the Weierstrass extreme value theorem. 

Next, noticing that $U_d(\theta) > 0 \in (0,1)$ we can derive that $\frac{\partial U_d}{\partial \theta} (0) > 0$ and $\frac{\partial U_d}{\partial \theta} (0) < 0$. Then if it only has one root in $(0,1)$, we would know $U_d$ must first increase and then decrease because there is at most one inflection point. Thus, a unique maximum exists.

\subsubsection{Proofs of why Uniform distribution has a unique maximized $\theta^{*}$}

If $\frac{\partial U_d}{\partial \theta}$ only has one root. We know it is first larger than 0, then becomes smaller than 0. Next, according to the Leibniz integral rule, we can get:

$$\frac{\partial U_d}{\partial \theta} = \int_{x^*(\theta)}^{\theta}P(x)dx - (\theta - x^*(\theta))\cdot P(x^*(\theta)) \cdot \frac{\partial x^*(\theta)}{\partial \theta}$$ 

Use Lagrange's Mean Value Theorem, we can write the above equation as: 

$$(\theta - x^*(\theta))\cdot [P(\theta^{'}) - P(x^*(\theta)) \cdot \frac{\partial x^*(\theta)}{\partial \theta}]$$

where $\theta^{'}$ is between $x^*(\theta), \theta$. Thus, the second term $P(\theta^{'}) - P(x^*(\theta)) \cdot \frac{\partial x^*(\theta)}{\partial \theta}$ must also be first larger than 0 then smaller than 0. Next, noticing that $P(\theta^{'}) = P(x^*(\theta))$ in uniform distribution and $\frac{\partial x^*(\theta)}{\partial \theta}$ is increasing, the equation will be smaller than 0 when $\frac{\partial x^*(\theta)}{\partial \theta} < 1$ and vice versa. Thus, we prove the result for the uniform distribution.

\subsubsection{Proof of Corollary \ref{corollary: theta change vs r}}

We now know $\frac{\partial U_d}{\partial \theta} = (\theta - x^*(\theta))\cdot [P(\theta^{'}) - P(x^*(\theta))] \cdot \frac{\partial x^*(\theta)}{\partial \theta}]$. Then according to the expression of $x^*(\theta)$, it is true that both $x^*(\theta)$ and $\frac{\partial x^*(\theta)}{\partial \theta}$ increase with $r$. Thus, when the probability distribution remains unchanged, the root of $\frac{\partial U_d}{\partial \theta}$ when $r$ increases becomes smaller.

\subsection{Proof Details of Thm. \ref{theorem: choice}}\label{subsection: proof2}

Denote $\ln\left(\sqrt{\frac{\theta^{-2}-1}{x_0^{-2}-1}}\right)$ as $G(x_0)$. $G(x_0)$ is always negative and monotonically increasing with $x_0 \in (0, \theta)$.

\paragraph{1. Situation when P = 0:}

According to Sec. \ref{sec:improvement} and \eqref{eq: utilde}, we can write the maximum improvement utility $U^*$ as $(1+r)^{\frac{G(x_0)}{ln(k^*+1)}} - k^*$, and write manipulation utility $\widetilde{U}$ as $(1+r)^{\frac{G(x_0)}{ln2}} - (\theta - x_0)$. 

Then take the derivative of both:
\abovedisplayskip=5pt
\begin{eqnarray}\label{eq:uhatstar}
\frac{\partial U^*}{\partial x_0} \ge \frac{\partial G}{\partial x_0}\cdot\frac{ln(1+r)}{ln(k^*+1)}\cdot(1+r)^{\frac{G(x_0)}{ln(k^*+1)}}
\end{eqnarray}

\begin{eqnarray}\label{eq:utildederivative}
\frac{\partial \widetilde{U}}{\partial x_0} = \frac{\partial G}{\partial x_0}\cdot\frac{ln(1+r)}{ln2}\cdot(1+r)^{\frac{G(x_0)}{ln2}} + 1
\end{eqnarray}

The ``$\ge$" in \eqref{eq:uhatstar} occurs because $k^*$ is actually a function of $x_0$, but if we regard $k^*$ at $x_0$ as a constant, the derivative here serves as a lower bound of $\frac{\partial U^*}{\partial x_0}$.

Firstly, we prove when $x_0 \rightarrow \theta$, $U^* < \widetilde{U}$: when $x_0 \rightarrow \theta$, we know $k^* \rightarrow 0$ since individuals invest an arbitrarily small effort to immediately qualified. However, according to Sec. \ref{subsection: proof1}, $k^*$ should let $\frac{\partial^2 U}{\partial k^2} < 0$. This inequality will give us the bound of $k^*$: $\ln(k^*+1) > \frac{-G(x_0)\cdot ln(1+r)}{3}$. With this bound, we can plug $k^*$ into \eqref{eq:uhatstar}, and know $\frac{ln(1+r)}{ln(k^*+1)} \rightarrow +\infty$, and $(1+r)^{\frac{G(x_0)}{ln(k^*+1)}}$ is larger than a constant because of the bound. Therefore, $\frac{\partial U^*}{\partial x_0} \geq \frac{\partial G}{\partial x_0}\cdot +\infty$. Then according to \eqref{eq:utildederivative}, when $x_0 \rightarrow \theta$, $\frac{\partial \widetilde{U}}{\partial x_0} < \frac{\partial G}{\partial x_0}\cdot\frac{ln(1+r)}{ln2} + 1$. Since $\frac{\partial G}{\partial x_0}$ is always positive, when $x_0 \rightarrow \theta$, we prove that $\frac{\partial U^*}{\partial x_0}>\frac{\partial \widetilde{U}}{\partial x_0}$. Meanwhile, when $x_0 = \theta$, $U^* = \widetilde{U} = 1$. This means when $x_0 \rightarrow \theta$, $U^* < \widetilde{U}$.

Secondly, when $x_0 = 0$: $\widetilde{U} = -\theta$ and $U^* = 0$. So $\widetilde{U} < U^*$ when $x_0 = 0$. 

Thus. there must be an intersection between $\widetilde{U}$ and $U^*$. Then noticing that if we increase $\theta$, $\widetilde{U}$ is always decreasing to converge to function $y = x - 1$, while $U^* \ge 0$ always holds. This suggests when $\theta$ is sufficiently close to 1, i.e., there exists a $\bar{\theta}$ and when $\theta > \bar{\theta}$, we can guarantee the first intersection of $U^*$ and $\widetilde{U}$ occurs arbitrarily close to 1, meaning this first intersection is the only intersection.

\begin{figure}
    \centering
    \includegraphics[width=0.3\linewidth]{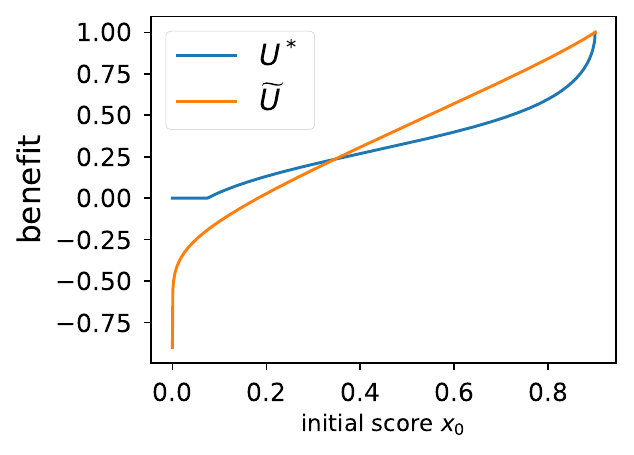}
    \caption{$\widetilde{U},U^*$}
    \label{fig:separate}
\end{figure}

Let the only \textit{intersection} be $\hat{x}$, we prove situation 1. The shapes of $\widetilde{U}$ and $U^*$ are illustrated in Fig. \ref{fig:separate}.

\paragraph{2. Situation when P $>$ 0:}

From \eqref{eq: utilde}: when $x_0 \rightarrow \theta$, $\widetilde{U} \rightarrow 1-P$. However, at this time $U^* \rightarrow 1 > 1-P$. This demonstrates $\hat{x_2}$ must exist.

When $P \rightarrow 0$, according to situation 1 and the continuity of $\widetilde{U}$ with respect to $P$, $\hat{x_1}$ must exist. However, when $P \rightarrow 1$, $\widetilde{U}$ is always negative, making $\hat{x_1}$ does not exist.

Thus, there must exist a threshold $\hat{P}$, when $P \le \hat{P}$, $\hat{x_1}, \hat{x_2}$ exist. Otherwise, $U^* > \widetilde{U}$ is always true.


\begin{figure}[tb]
     \centering
    \begin{subfigure}[b]{0.25\textwidth}
         \includegraphics[width=\textwidth, height = 100pt]{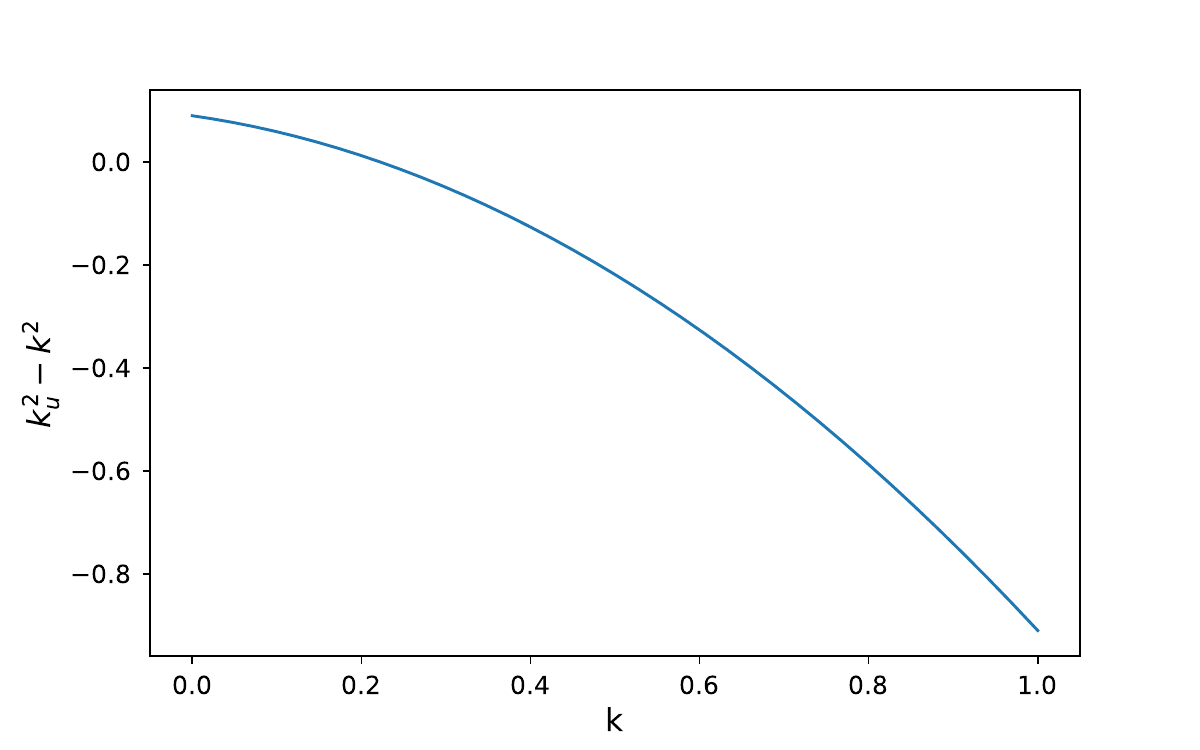}
         \caption{shape when $2x_0^2+2x_0^3-1 < 0$}
     \end{subfigure}
     \hspace{1em}
     \begin{subfigure}[b]{0.25\textwidth}
         \includegraphics[width=\textwidth, height = 100pt]{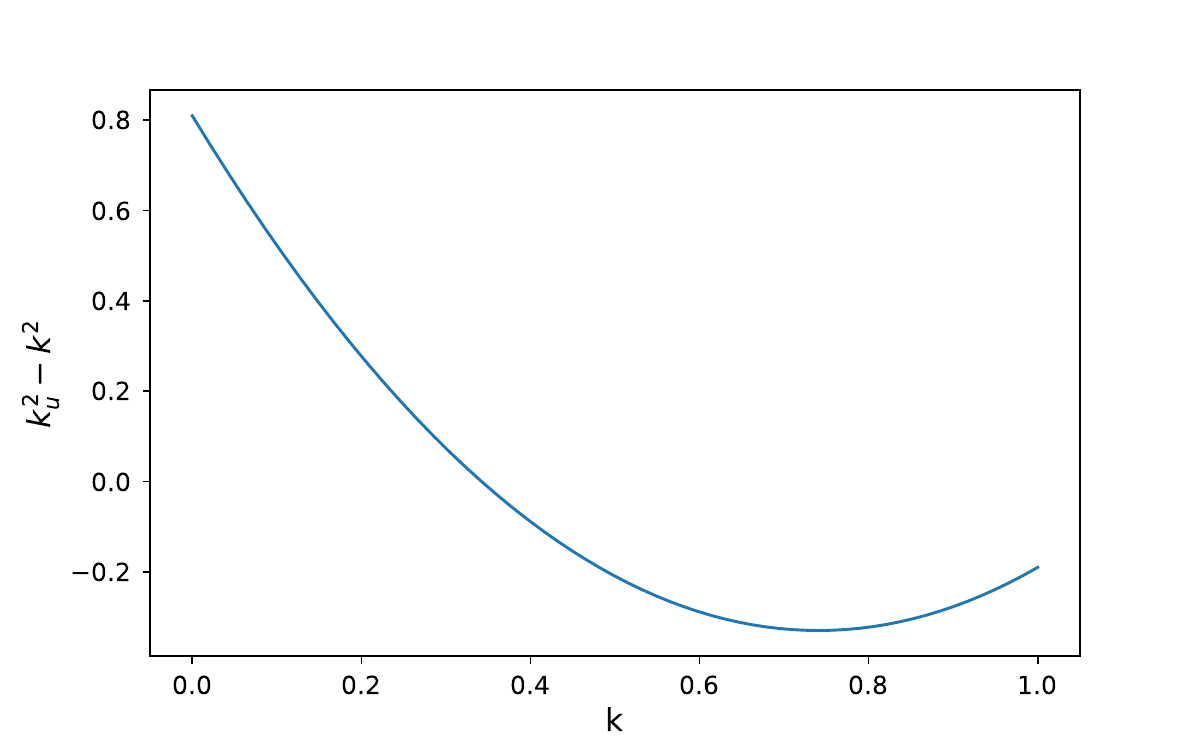}
         \caption{shape when $2x_0^2+2x_0^3-1 > 0$}
     \end{subfigure}
     \caption{Shapes of $k_u^2-k^2$}
\label{fig:illu6}
\end{figure}

\subsection{Proof Details of Thm. \ref{theorem: fconverge}}\label{subsection: proof3.0}

Thm. \ref{theorem: fconverge} can be proved by the inequality $\|\widetilde{d}\|q_t^Td > \|\widetilde{d}\|q_0^Td > \|\widetilde{d}\|q_0^Td q_t^Td^{*}$. This means we can just let $k_u = \|\widetilde{d}\|q_0^Td = \|\widetilde{d}\|x_0$ and directly apply Lemma \ref{lemma:convergence}.

\subsection{Proof Details of Thm. \ref{theorem: fimprove}}\label{subsection: proof3}

First let us prove following two lemmas:

\begin{lemma}\label{lemma: khat}
For any initial qualification score $x_0$, There exists a $\widehat{k} \in (0,1)$, when $k \in [0,\widehat{k}$), $k_u > k$. Let $\widehat{x_0}$ be the only root of $2x_0^2+2x_0^3-1 = 0$ within $(0,1)$, then $\widehat{k}$ is given by:
\small
\begin{eqnarray}\label{eq:khatdef2}
\widehat{k} = \min(\frac{\widehat{x_0}^2}{2\widehat{x_0}^2 + 2\widehat{x_0}^3}, \frac{x_0\cdot(x_0^2+x_0 - \sqrt{x_0^4-x_0^2+1})}{2x_0^2+2x_0^3-1})
\end{eqnarray}
\normalsize
\end{lemma}

\paragraph{Proof.}

According to Thm. \ref{theorem: fconverge}, $k_u^2 = \|\widetilde{d}\|^2\cdot x_0^2$ and $\|\widetilde{d}\|^2 = k^2+(1-k)^2-2k(1-k)x_0$. We can get following expression:

\begin{eqnarray}\label{eq:kumk}
k_u^2 - k^2 = (2x_0^2+2x_0^3-1)k^2 - (2x_0^2 + 2x_0^3)k + x_0^2
\end{eqnarray}

Firstly, when $2x_0^2+2x_0^3-1 = 0$, $\widehat{x_0}= 0.565$. Thus, when $k < \frac{\widehat{x_0}^2}{2\widehat{x_0}^2 + 2\widehat{x_0}^3} = 0.319$, $k_u^2 > k^2$.

Except the above situation, We can regard \eqref{eq:kumk} as a quadratic function of $k$ and solve the two roots:

\begin{eqnarray}\label{eq:root21}
\frac{x_0\cdot(x_0^2+x_0 \pm \sqrt{x_0^4-x_0^2+1})}{2x_0^2+2x_0^3-1}
\end{eqnarray}

We then prove a claim that when $x_0 \in (0,1)$, $\frac{x_0\cdot(x_0^2+x_0 + \sqrt{x_0^4-x_0^2+1})}{2x_0^2+2x_0^3-1}$ is either larger than 1 or smaller than 0:

1) When $2x_0^2+2x_0^3-1 < 0$, the denominator of \eqref{eq:root21} is negative, while the numerator is always positive. Thus, \eqref{eq:root21} is negative.

2) When $2x_0^2+2x_0^3-1 > 0$:

\begin{eqnarray}\label{eq:root22}
\frac{x_0\cdot(x_0^2+x_0+\sqrt{x_0^4-x_0^2+1})}{2x_0^2+2x_0^3-1} > \frac{x_0\cdot(x_0^2+x_0+x_0^2)}{2x_0^3+x_0^2} = 1
\end{eqnarray}

\eqref{eq:root22} means \eqref{eq:root21} is larger than 1. Thus, the claim is proved.

Thus, $k_u^2 - k^2$ only has one root within $(0,1)$. Also from \eqref{eq:kumk} we know when $k = 0$, $k_u > k$ and when $k = 1$, $k_u \le k$. With these facts we immediately know: When $k \le\frac{x_0\cdot(x_0^2+x_0 - \sqrt{x_0^4-x_0^2+1})}{2x_0^2+2x_0^3-1}$, $k_u^2 - k^2 \ge 0$. Otherwise, $k_u^2 - k^2 < 0$. In fact, besides the exception $2x_0^2+2x_0^3-1 = 0$, there are only two possibilities of the shape of $k_u^2 - k^2$ as shown in Fig. \ref{fig:illu6}. Because $k$ and $k_u$ are both non-negative, the relationship of the square must be the same for their values.

Then if we define $\hat{k}$ as:

\begin{eqnarray}\label{eq:kfinal}
\hat{k} = \min(\frac{\widehat{x_0}^2}{2\widehat{x_0}^2 + 2\widehat{x_0}^3}, \frac{x_0\cdot(x_0^2+x_0 - \sqrt{x_0^4-x_0^2+1})}{2x_0^2+2x_0^3-1})
\end{eqnarray}

Then $k_u > k$ when $k \in [0,\hat{k})$. Proved.

\begin{lemma}\label{lemma: khat2}
For any individual with initial qualification score $x_0$ and the admission threshold $\theta$, there must exist a $r$ to let there exists a $\bar{k} \in [0,\widehat{k})$, 
$U(\bar{k}, \theta, r, x_0) > 0$
\end{lemma}

\paragraph{Proof.} If we let $z = \ln(k+1)$ be $z$ and recall that $C(\theta, x_0, r) = -\ln\left(\sqrt{\frac{(\theta)^{-2}-1}{(x_0)^{-2}-1}}\right)\cdot\ln(1+r)$, we would have $U = e^{\frac{-C}{z}} - e^{z} + 1$.

For any $z$ there exists $C_z$, when $C < C_z$, $U > 0$. 

So we can just let $k$ be an arbitrary point $\in [0,\widehat{k})$ and we can get the corresponding $C_z$, then we can only let $r$ satisfy:

\begin{eqnarray}\label{conditionr}
\ln(1+r) < \frac{C_z}{-\ln\left(\sqrt{\frac{(\theta)^{-2}-1}{(x_0)^{-2}-1}}\right)}
\end{eqnarray}

Then we find the plausible $r$. Proved.

\paragraph{Proof of Thm. \ref{theorem: fimprove}.} According to Lemma \ref{lemma: khat}, when $\bar{k} \in [0,\widehat{k})$, $k_u > k$, so the convergence speed of the individual to $d^*$ under forgetting mechanism will be faster than the convergence speed of the individual to $d$ without forgetting mechanism, so that the reward under forgetting mechanism is discounting less. Meanwhile, according to Lemma \ref{lemma: khat2}, there exists a $r$ where $U(\bar{k},\theta, r, x_0) > 0$. Combine them together, $\widehat{U}(\bar{k}, \theta, r, x_0) > U(\bar{k},\theta, r, x_0) > 0$ and Thm. \ref{theorem: fimprove} is proved.

\subsection{Proof of Thm. \ref{theorem:generalized}}\label{subsection:proofg}

Assume $k_t = \frac{k}{t+1}$ when $t \ge 0$. From \eqref{eq:dynamics} and similar to \cite{Dean2022}, we know $(q_{t+1}^T\cdot d)^{-2} - 1 = \frac{(q_{t}^T\cdot d)^{-2} - 1}{k_t+1}^2$. This will lead to $(q_{t}^T\cdot d)^{-2} - 1 = \prod_{i=0}^{t-1}(\frac{k}{i+1} + 1)^{-2} ((q_{0}^T\cdot d)^{-2} - 1)$.

Then consider $\prod_{i=0}^{t-1}(\frac{k}{i+1} + 1)^{-1} = \prod_{i=0}^{t-1}(\frac{i+1}{k+i+1}) = \frac{1}{k+1} \cdot \frac{2}{k+2} ...$. When $k = 1$, The expression is $\frac{1}{2}\cdot \frac{2}{3} \cdot \frac{3}{4}...\frac{t-1}{t} = \frac{1}{t}$, demonstrating the convergence rate is linear. Note that this expression is decreasing as $k$ decreases, so the convergence rate in our model is always slower than linear. Next, consider the general expression $\prod_{i=0}^{t-1}(\frac{i+1}{k+i+1}) = \frac{1}{k+1} \cdot \frac{2}{k+2} ...$ and $k < 1$. Let $a = \frac{1}{k}$ which is larger than $1$, and $j = i+1$ which is larger than $0$. We slightly abuse the definition of $a$ to let it be an integer. Then the expression becomes $\prod_{i=0}^{t-1}(\frac{ja}{1+ja}) = \frac{a}{a+1} \cdot \frac{2a}{2a+1} ... \frac{ta}{ta+1}$. 

Then for any $a$ we can bound this expression. Basically, we already know $\frac{1}{2}\cdot \frac{2}{3} \cdot \frac{3}{4}...\frac{t-1}{t} = \frac{1}{t}$. Noticing that when $a > 1$, it is just equal to erase some terms of this expression. We can utilize this fact to get the lower bound and upper bound:

\begin{enumerate}
    \item Lower bound: consider the following $a-1$ sets of expressions and each set consists of $t$ terms: $\{\frac{1}{2} \cdot \frac{a+1}{a+2} \cdot \frac{2a+1}{2a+2} ... \cdot \frac{(t-1)a+1}{(t-1)a+2}\}$, $\{\frac{2}{3} \cdot \frac{a+2}{a+3} \cdot \frac{2a+2}{2a+3} ... \cdot \frac{(t-1)a+2}{(t-1)a+3}\}$..., $\{\frac{a-1}{a} \cdot \frac{2a-1}{2a} \cdot \frac{3a-1}{3a} ... \cdot \frac{ta-1}{ta}\}$. Then each of the $a-1$ expressions are smaller than $\prod_{j=1}^{t}(\frac{ja}{1+ja}) = \frac{a}{a+1} \cdot \frac{2a}{2a+1} ... \frac{ta}{ta+1}$. Denote $\prod_{j=1}^{t}(\frac{ja}{1+ja}) = \frac{a}{a+1} \cdot \frac{2a}{2a+1} ... \frac{ta}{ta+1}$ as $I$, we will have $I^{a} \ge \frac{1}{ta+1}$, so the convergence rate is smaller than $\sqrt[a]{ta} = \Theta(t^k)$

    \item Upper bound: consider the following $a-1$ sets of expressions and each set consists of $t$ terms: $\{\frac{a+1}{a+2} \cdot \frac{2a+1}{2a+2} ... \cdot \frac{ta+1}{ta+2}\}$, $\{\frac{a+2}{a+3} \cdot \frac{2a+2}{2a+3} ... \cdot \frac{ta+2}{ta+3}\}$..., $\{\frac{2a-1}{2a} \cdot \frac{3a-1}{3a} ... \cdot \frac{(t+1)a-1}{(t+1)a}\}$. Then each of the $a-1$ expressions are larger than $\prod_{j=1}^{t}(\frac{ja}{1+ja}) = \frac{a}{a+1} \cdot \frac{2a}{2a+1} ... \frac{ta}{ta+1}$. Denote $\prod_{j=1}^{t}(\frac{ja}{1+ja}) = \frac{a}{a+1} \cdot \frac{2a}{2a+1} ... \frac{ta}{ta+1}$ as $I$, we will have $I^{a} \le \frac{1}{(t+1)}$, so the convergence rate is larger than $\sqrt[a]{ta} = \Theta(t^k)$
\end{enumerate}

Thus, take the limit and apply the Sandwich Theorem, the convergence rate is $\Theta(t^k)$.

\end{document}